\documentclass[]{interact}
\usepackage{color}
\usepackage{caption}
\usepackage{graphicx}
\usepackage[version=4]{mhchem}
\usepackage{cases}
\usepackage{enumerate}
\usepackage{amsmath}
\usepackage{epstopdf}
\usepackage[caption=false]{subfig}

\usepackage[numbers,sort&compress]{natbib}
\bibpunct[, ]{[}{]}{,}{n}{,}{,}
\makeatletter
\def\NAT@def@citea{\def\@citea{\NAT@separator}}
\makeatother

\theoremstyle{plain}
\newtheorem{theorem}{Theorem}[section]

\theoremstyle{definition}

\theoremstyle{remark}
\newtheorem{remark}{Remark}

\begin{document}


\title{Recent Advances in Conservation-Dissipation Formalism for Irreversible Processes} 

\author{
\name{Liangrong Peng\textsuperscript{a} and Liu Hong\textsuperscript{b}\footnote{Author to whom correspondence should be addressed: penglr.wh@qq.com (LRP) hongliu@sysu.edu.cn (LH)}}
\affil{\textsuperscript{a}College of Mathematics and Data Science, Minjiang University, Fuzhou, 350108, P.R.C. \\
\textsuperscript{b}School of Mathematics, Sun Yat-sen University, Guangzhou, 510275, P.R.C.}
}

\maketitle

\begin{abstract}
The main purpose of this review is to summarize the recent advances of the Conservation-Dissipation Formalism (CDF), a new way for constructing both thermodynamically compatible and mathematically stable and well-posed models for irreversible processes. The contents include but are not restricted to the CDF's physical motivations, mathematical foundations, formulations of several classical models in mathematical physics from master equations and Fokker-Planck equations to Boltzmann equations and quasi-linear Maxwell equations, as well as novel applications in the fields of non-Fourier heat conduction, non-Newtonian viscoelastic fluids, wave propagation/transportation in geophysics and neural science, soft matter physics, \textit{etc.} Connections with other popular theories in the field of non-equilibrium thermodynamics are examined too.
\end{abstract}


\tableofcontents

\section{Introduction}
The last half century has witnessed a rapid progress in non-equilibrium thermodynamics, which becomes an exciting and fruitful research field in modern physics.
Non-equilibrium thermodynamics abandons several ideal assumptions of the equilibrium approach and leads to much broader and realistic studies beyond equilibrium. Moreover, it provides a powerful and unified framework to handle various irreversible processes arising from physics, chemistry, biology, engineering, and so on. Based on different standpoints and assumptions, there formed many ``schools" of non-equilibrium thermodynamics during the past years.

The beginning of modern non-equilibrium thermodynamics is usually dated back to 1931, in which year Onsager established his well-known ``reciprocal relations" based on the time reversibility of microscopic dynamics and linear regression hypothesis. Later, together with the pioneering works of Prigogine, de Groot, Mazur and many others, classical irreversible thermodynamics \cite{de2013} (CIT) was developed into the first systematical theory for non-equilibrium thermodynamics. Based on the local-equilibrium hypothesis, which assumes a system locally is always in equilibrium but globally can still be varying slowly, CIT finds its applications in a wide range of scientific and industrial areas. Unfortunately, the local-equilibrium hypothesis, and thus CIT fails when short time/space scales are involved.

Later, constructed on several basic axioms, including the principles of material frame-indifference, fading memory, local action and equipresence, Coleman \cite{Coleman1964Thermodynamics}, Truesdell \cite{Truesdell1984Rational}, Noll \cite{Noll1974The} \textit{et al.} developed the rational thermodynamics (RT). RT provides valuable insights into the modeling and receives plenty of successes in the field of complex fluids. The main criticisms on RT concern about the unclear definition of temperature and entropy, the complexity of constitutive equations, \textit{etc.} \cite{jou1999}

To overcome the drawbacks of local-equilibrium hypothesis and linear constitutive relations, extended irreversible thermodynamics (EIT) was proposed by M{\"u}ller and Ruggeri \cite{muller2013extended}, Jou, Casas-V{\'a}zquez and Lebon \cite{jou1996extended}, \textit{etc}.
Besides the classical conserved variables in CIT (such as mass, momentum, energy), EIT enlarges the space of independent state variables by including the dissipative ones too (such as stress tensor, heat flux, \textit{etc.}). EIT is suitable for characterizing systems with short relaxation times, for instance the flow of polymeric fluids and heat transport in nano-systems. However, the mathematical foundation of EIT has never been rigorously justified. Brutal inclusion of dissipative fluxes into the space of state variables sometimes fails too.

Following a distinct routine, Grmela \cite{Grmela1989Hamiltonian} proposed a Hamiltonian version of non-equilibrium thermodynamic theory, which is subsequently developed into the general equation for the non-equilibrium reversible-irreversible coupling (GENERIC) form by Grmela and {\"O}ttinger \cite{Grmela1997Dynamics,Ottinger1997Dynamics}. GENERIC could be considered as a direct generalization of Hamiltonian equations for conserved dynamics and Ginzburg-Landau equations for dissipative dynamics, which possesses the structure of contact geometry. Although the Poission brackets are somehow hard to determine, and no mature numerical algorithms are available yet, GENERIC has been widely applied to rheology and polymeric fluids. By selecting the conjugate dual of flux as state variables, Lebon \textit{et al.} \cite{lebon2017extended} proposed a compatible formalism to link both EIT and GENERIC.

Physicists and engineers prefer to do the modeling through the variational approach, which enjoys a clear physical interpretation. For instance, the Lagrangian dynamics in classical mechanics can be deduced based on the least action principle. In the presence of friction, Rayleigh generalized the Lagrangian equation by adding an extra dissipative potential into the action.
More recently, Doi followed Rayleigh's idea and developed a variational approach by incorporating Onsager's reciprocal relations. The phenomenological equations derived in this way essentially show that, the time evolution of a system is determined by the balance between potential forces and frictional forces. The potential force drives the system into a state of potential minimum, while the frictional force resists the trend. It turns out that Doi's variational principle is valid for many problems in soft matter physics \cite{doi2013soft}, from thin films \cite{Xu2015A}, viscoelastic filaments \cite{Zhou2018Dynamics} and solid toroidal islands \cite{Jiang2019Application}, to the deposition patterns of two droplets next to each other \cite{Hu2017Deposition} and boundary conditions for liquid-vapor flows and immiscible two-phase flows \cite{Xu2017Hydrodynamic}.

On the other hand, Liu \textit{et al.} proposed an energetic variational approach (EVA), which focuses on the coupling between fluid flows and internal micro-structures of particles. The least action principle gives the Hamiltonian part for hydrodynamics, while the maximal dissipation principle deduces the Onsager's part.
EVA provides a self-consistent method to study complex fluids, for instance, vesicles interacting with fluids \cite{Du2009Energetic}, flows of nematic liquid crystals \cite{Sun2009On}, two-phase fluids \cite{Hyon2017Energetic}, non-isothermal electrokinetics and so on \cite{liu2017non}.


Established on the modern theory of first-order hyperbolic equations, conservation-dissipation formalism of irreversible thermodynamics (CDF) can be seen as a mathematically regularized theory of EIT and GENERIC.
CDF is rooted in the generalized nonlinear version of Onsager's reciprocal relations \cite{yong2008, Peng2018Generalized} and the Godunov structure for symmetrizable hyperbolic equations \cite{godunov1961interesting, friedrichs1971systems}, which in turn guarantees the hyperbolicity of models, well-posedness and globally asymptotic stability of solutions, as well as well-behaved limits of corresponding relaxation problems \cite{zhu2015conservation}. As a rigorous formalism in mathematics, CDF has been applied to plenty of non-equilibrium systems, \textit{e.g.}, non-Fourier and non-ballistic heat conduction in nano-scales \cite{huo17, jozsa2020general}, isothermal and non-isothermal flows of compressible viscoelastic fluids \cite{zhu2015conservation, huo2016structural, Yang2018Generalized}, wave propagation in saturated porous media \cite{liu2016stability, Liu2016Stability2}, axonal transport with chemical reactions \cite{Yan2011Weak, yan2012stability}, and so on. Additionally, an interesting connection with moment hierarchies of the Boltzmann equation was established \cite{hong2015novel}, which puts CDF on a solid mesoscopic kinetic foundation. Recently, CDF has gained wide attention in the field of non-equilibrium thermodynamics. Its connection and distinction with other ``schools'', including the aforementioned EIT \cite{sun2016nonlinear, lebon2017extended}, GENERIC \cite{lebon2017extended} and steady-state thermodynamics for mesoscopic stochastic processes \cite{qian2016entropy}, \textit{etc}., were discussed in detail from time to time.

This paper aims to present a comprehensive review on the mathematical foundation and physical motivations of CDF, to summarize all well-known classical models in mathematical physics which fall into the category of CDF, and to show its novel applications in various disciplines. The whole paper is organized as follows.
Sec. \ref{general formulation} is devoted to the mathematical foundations and physical motivations of CDF, with emphasis on recipes for CDF modeling illustrated through the generalized Navier-Stokes-Fourier equations.
As validations of CDF, in Sec. \ref{applications I} we reformulate some well-known models in mathematical physics into the CDF structure, such as the Boltzmann equation, master equation, Fokker-Planck equation, mass-action equation, \textit{etc.}
Furthermore, we apply CDF to derive unknown constitutive relations in various fields, including viscoelastic fluids, heat conduction, soft matter physics, geophysics, and so on. This part of results is summarized in Sec. \ref{applications II}. In Sec. \ref{validation}, the advantages of CDF over other physical approaches are demonstrated from three aspects -- mathematical analysis, numerical simulations and experimental validations based on recent advances. The last section is a conclusion.

\section{Physical Motivation and Mathematical Foundation}\label{general formulation}

\subsection{Symmetry, scale separation and conservation laws}

Instead of creating from nothing, we refer to the conservation laws as our starting point. The significance of conservation laws in mathematics and physics can not be overemphasized. In fact, the classical mechanics and thermodynamics in the 19th century and before can be regarded as a history of discovery of momentum and energy conservation in some sense. For example, through the Hamiltonian equations, the classical non-dissipative mechanics becomes a subject on how to construct a concrete form of the Hamiltonian -- the energy function and how to solve the Hamiltonian equations explicitly or implicitly. A similar conclusion also holds for quantum mechanics, using the Schr{\"o}dinger equation instead.

Why do conservation laws of mass, momentum and energy play such a key role in natural sciences? A most insightful answer was provided by Noether, a great mathematician, who stated that ``every differentiable symmetry of the action of a physical system has a corresponding conservation law''. As an illustration, if a physical system exhibits the same outcome regardless of how it is translated in space or time, then by Noether's theorem, these symmetries account for the conservation laws of linear momentum and energy respectively. As another example, if the behavior of a physical system does not change upon spatial or temporal reflection, the parity/entropy of the system will be conserved as a consequence of laws of motion. Noether's theorem provides a direct connection between the symmetry properties of a system and its conservation laws. The former is actually a geometrical argument, while the latter is analytic.

Noether's theorem is restricted to systems that can be modeled with a Lagrangian alone. In particular, dissipative systems with continuous symmetries do not need to have a corresponding conservation law. However, if a system involves processes happening in multiple time/space scales, the conservation laws can be recovered to a certain degree. To see this point, let us take the fast-slow dynamics as an illustration. Consider the enzyme catalyzed reaction $$S+E\rightleftharpoons ES\rightarrow P+E,$$ where $E, S, P, ES$ denote the enzyme, substrate, product and intermediate complex respectively. Clearly, there are only two independent conservation laws relating with the mass concentrations of enzyme and substrate, \textit{i.e.},  $[E]+[ES]=const, [S]+[ES]+[P]=const$. If we further assume the conversion from $ES$ to $E+P$ is much slower than the binding and unbinding reactions between enzyme and substrate, according to the famous Michaelis-Menten mechanism, there will be a ``perfect'' equilibrization among $E$, $S$ and $ES$. As a consequence, an extra conservation law $[S][E]/[ES]=const$ emerges in the time scale of slow dynamics. Actually, this is also the key point of partial equilibrium approximation. In mathematics, the procedure of neglecting the fast dynamics and only focusing on the slow dynamics is a projection of high-dimensional trajectories onto a lower-dimensional manifold. And the restrictions are expressed through new conservation laws.

\subsection{Entropy, free energy and Onsager's relation}

Conservation laws highlights the reversible aspect of a process, while its irreversible aspect is characterized by the entropy function and its time derivatives. Just as stated by the first law and the second law of thermodynamics, the total energy of an isolated system is constant, which can be transformed from one form to another, but never be created nor destroyed. In contrast, entropy is a monotonically increasing function with its maximum obtained at the equilibrium state. Therefore, energy and entropy  act as two sides of the same coin. Together they constitute the whole story of a non-equilibrium process.

Entropy is one of the most mysterious and most controversial concepts in physics. From a thermodynamic point of view, it is closely related to the dissipative heat, or the irreversibility of how useful work is converted into non-useful heat and vice versa, as stated either in Carnot cycle or in Clausius inequality. While from a statistical mechanical point of view, entropy is a counting or statistics of all possible microscopic configurations/states of a given system in the equilibrium. To extend the concept of equilibrium entropy to general non-equilibrium states, Boltzmann took the first significant step forward by introducing the famous H-function for the Bolzmann equation. The H-function, defined as the ensemble average of the logarithm of the instantaneous distribution function, is not restricted to equilibrium states anymore. It enjoys an elegant property of monotonically decreasing in time and thus serves as the Lyapunov function for the Boltzmann function. The idea of H-function was later generalized to the Boltzmann-Gibbs entropy in probability, the Shannon entropy in information theory, the Tsallis or Renyi entropy in non-extensive statistical physics, the trajectory entropy in stochastic process, and so on.

Through Legendre transformations, the extremal requirement on the Boltzmann-Gibbs-Shannon entropy can be transformed into that on the Helmholtz free energy or Gibbs free energy. The free energy, or relative entropy, by making use of a preknowledge of the equilibrium or steady-state distribution, is a more suitable quantity for characterizing the irreversibility of a non-equilibrium process in mathematics, just as stated through the famous Kullback-Leibler divergence. Recently, the large deviations principle for a given stochastic process provides a systematic way to derive the free energy function for both CIT and EIT \cite{Hong20-1, Hong20-2}.

The non-negativity of the entropy production rate is an alternative statement of the second law of thermodynamics. Onsager made an astonishing observation by writing the entropy production rate into a bilinear product of thermodynamic forces and fluxes, which are connected to each other through the dissipation matrix. Especially in the linear region not far away from equilibrium, it has been shown that the dissipation matrix is constant, non-negative and symmetric, which is known as Onsager's reciprocal relations in literature. The Newton's law for viscosity, Fourier's law for heat conduction, Ohm's law for electricity, \textit{etc.} are all manifestations of Onsager's relation in different fields. In classical irreversible thermodynamics, the semi-positive entropy production rate as well as Onsager's relation constitutes two golden criterions for modeling, analysis and applications.

\subsection{The conservation-dissipation formalism} \label{The CDF}
In this part, we would like to present the general formulation of CDF, which consists of four key steps as illustrated in what follows.

Firstly, \textbf{to choose suitable conserved and dissipative variables}. In CDF, not only classical conserved variables, like mass, momentum and energy, which are widely adopted in classical continuum mechanics and hydrodynamics, but also dissipative variables related to the irreversibility of non-equilibrium processes, are required to provide a comprehensive description of the system. It is directly shown that conserved variables obey some conservation laws expressed as
\begin{equation}\label{con}
  \partial_t \textbf{y}+\sum_{j=1}^{\Lambda} \partial_{x_j}J_j=0,
\end{equation}
where ${\Lambda}$ is the dimension of space, the vector $\textbf{y}=\textbf{y}(t,\textbf{x}) \in \mathbb{R}^n$ depends on $t\in \mathbb{R}^1_{\geq 0}$ and $\textbf{x} \in \mathbb{R}^{\Lambda}$. $J_j$ is the flux in the $x_j$ direction.
Once the form of flux $\textbf{J}=\textbf{J}(\textbf{y},\cdots)$ is specified, we would get a closed form of partial differential equations. And thus the system dynamics is completely determined provided suitable initial and boundary conditions.

Towards dissipative variables, it is noted that EIT includes unknown fluxes appearing in the conservation laws directly (such as heat flux, stress tensor), while CDF suggests to adopt the conjugate variables of fluxes with respect to the entropy function \cite{zhu2015conservation}. That is, the dissipative variables $\textbf{z}=\textbf{z}(t,\textbf{x}) \in \mathbb{R}^m$ are specified such that the flux is expressed as $J_j=\partial s(\textbf{y},\textbf{z})/\partial z_j$. The time change of $\textbf{z}$ is assumed to satisfy the balance equations,
\begin{equation}\label{diss}
  \partial_t \textbf{z}+\sum_{j=1}^{\Lambda} \partial_{x_j}K_j(\textbf{y},\textbf{z})=\textbf{q}(\textbf{y},\textbf{z}),
\end{equation}
where $K_j(\textbf{y},\textbf{z})$ denotes the flux corresponding to dissipative variables $\textbf{z}$ in the $x_j$ direction, $\textbf{q}(\textbf{y},\textbf{z})$ is the source term and vanishes when the system is at equilibrium.
For notational convenience, we rewrite the conservation and balance laws \eqref{con}-\eqref{diss} together as
\begin{equation}\label{con-diss}
  \partial_t \textbf{U}+\sum_{j=1}^{\Lambda} \partial_{x_j}F_j(\textbf{U})=\textbf{Q}(\textbf{U}),
\end{equation}
where
\begin{align*}
\textbf{U}=\begin{pmatrix}\textbf{y}\\\textbf{z} \end{pmatrix}, \quad F_j(\textbf{U})=\begin{pmatrix}J_j(\textbf{U})\\K_j(\textbf{U}) \end{pmatrix}, \quad \textbf{Q}(\textbf{U})=\begin{pmatrix}0\\\textbf{q}(\textbf{U}) \end{pmatrix}.
\end{align*}
This is the fundamental form of CDF.

Remark that the adoption of conjugate variables $\textbf{z}$ rather than directly taking the flux $\textbf{J}$ as independent variables plays an essential role in CDF, and has a long history in equilibrium thermodynamics (\textit{e.g.}, Legendre transformations).
As Sun \textit{et al.} \cite{sun2016nonlinear} pointed out, by selecting the thermodynamic conjugate of an extra stress rather than stress itself, CDF provides a suitable framework for constructing genuinely nonlinear models for non-Newtonian fluids, while EIT fails to do this.

Secondly, \textbf{to construct a strictly concave entropy function $s=s(\textbf{y},\textbf{z})$}. There are several general comments on the entropy function. (1) The free energy or relative entropy is more proper than the entropy for isothermal systems. For instance, the classical entropy form used for the Boltzmann equation, master equation, Fokker-Planck equation and mass-action equations is the Boltzmann-Gibbs entropy $\int p \ln p d\textbf{x}$, while the corresponding free energy is $\int p \ln(p/p^s) d\textbf{x}$. Here $p$ is interpreted as the concentrations of species in the mass-action equations or the probability (density) for the others. The integration reduces to the summation of states in discrete cases. With the help of an extra preknowledge on the steady-state concentration or probability distribution, it can be rigorously proved that the free energy rather than the entropy will be monotonically decreasing in the examples we listed. (2) For general continuum mechanics, it is still an open problem to specify the concrete form of the entropy function. Without further physical insights of the system, one can always start with a non-equilibrium entropy as a summation of the equilibrium entropy and a quadratic function of dissipative variables, which is valid near equilibrium.

Thirdly, \textbf{to calculate the entropy flux and entropy production rate}. Thanks to the Gibbs relation, the entropy change can be split into a full divergent term and a non-negative term,
\begin{equation}\label{ent-diss}
  \partial_t s=-\sum_{j=1}^{\Lambda} s_{\textbf{U}}\cdot\partial_{x_j}F_j(\textbf{U})+s_{\textbf{U}}\cdot \textbf{Q}(\textbf{U})=-\sum_{j=1}^{\Lambda} \partial_{x_j}J^f_j(\textbf{U})+s_{\textbf{z}}(\textbf{U})\cdot \textbf{M}(\textbf{U}) \cdot s_{\textbf{z}}(\textbf{U}),
\end{equation}
which are recognized as the entropy flux and entropy production rate respectively. And we notice the second law of thermodynamics is automatically guaranteed with respect to the condition $\textbf{M} \geq 0$,
since the entropy production rate is always non-negative. Together with the first law of thermodynamics stated through energy conservation in (\ref{con}), we have constructed a mathematical formulation for modeling irreversible processes in compatible with general thermodynamic requirements.

Finally, \textbf{to specify the dissipation matrix $\textbf{M}=\textbf{M}(\textbf{y},\textbf{z})\geq0$}. During the procedure of reformulating entropy flux and entropy production rate, the unknown constitutive relations for dissipative variables $\textbf{z}$ in (\ref{diss}) will be totally specified. Now the only missing part is a concrete form of the dissipation matrix $\textbf{M}$. However, except for the positive semi-definite requirement, we nearly know nothing more about $\textbf{M}$ in general. It is not only material-dependent, but also process-dependent, which really brings big troubles to modeling. Currently, we still have to specify the dissipation matrix case by case, but we do know that they must be compatible with classical linear laws when the system is not far away from the local equilibrium, which means they are constant, positive and symmetric as claimed by Onsager's relation.

\subsection{Structural conditions for the existence of global smooth solutions}
In physical based modeling approaches, the existence of global smooth solutions is seldom considered. However, in mathematics the well-posedness of solutions is a key requisition for a successful model. In CDF, we try to make a nice balance between the physical meaning and mathematical rigorousness, which gives rise to the structural conditions of CDF.

First of all, in the absence of the source term $\textbf{Q}(\textbf{U})=0$, we notice the system in Eq. \eqref{con-diss} reduces to a system of local conservation laws. In that case, it is well-known that even when the initial data are smooth and close to equilibrium values, the solutions will generally develop singularities in finite time, \textit{e.g.} the shock waves and shock structure in gas dynamics \cite{dafermos2010hyperbolic}. As a result, we expect the source terms to prevent the breakdown of smooth solutions by imposing proper conditions.

Based on mathematical studies on hyperbolic relaxation problems, as well as the consistency with results obtained by CIT in local equilibrium, a group of structural conditions for CDF in Eq. \eqref{con-diss} are proposed \cite{Yong2004entropy}:
\begin{enumerate}[(a)]
\item there is a strictly concave smooth entropy function $s(\textbf{U})$ defined in a convex compact neighborhood $G$ of $\textbf{U}_e$, such that $s_{\textbf{UU}} \cdot F_{j\textbf{U}}$ is symmetric $\forall j=1,2,\cdots, \Lambda$ and $\forall \textbf{U}=(\textbf{y},\textbf{z})^T$ under consideration; \label{condition 1}
\item there is a dissipation matrix $\textbf{M}=\textbf{M}(\textbf{y},\textbf{z})\geq0$ such that $\textbf{q}(\textbf{U})=\textbf{M}(\textbf{U}) \cdot s_{\textbf{z}}(\textbf{U})$;
    \label{condition 2}
\item the kernel of $\textbf{Q}_\textbf{U}(\textbf{U}_e)$ contains no eigenvector of the matrix $\sum_j w_jF_{j\textbf{U}}(\textbf{U}_e)$ $\forall \textbf{w}=(w_1,\cdots, w_{\Lambda})\in \mathbb{S}^{{\Lambda}-1}$ (the unit sphere in $\mathbb{R}^{\Lambda}$);
    \label{condition 3}
\end{enumerate}
where $s_{\textbf{UU}}=\frac{\partial^2 s}{\partial \textbf{U}^2}, F_{j\textbf{U}}=\frac{\partial F_{j}}{\partial \textbf{U}}$, and $s_{\textbf{z}}=\frac{\partial s}{\partial \textbf{z}}$. $\textbf{U}_e$ is a constant vector representing the equilibrium state by satisfying the condition $\textbf{Q}(\textbf{U}_e)=0$.

The first condition (\ref{condition 1}) provides that the system \eqref{con-diss} is globally symmetrizable hyperbolic, which is known as the Lax entropy condition for hyperbolic conservation laws \cite{dafermos2010hyperbolic,friedrichs1971systems}. By the Poincare lemma, $s_{\textbf{UU}} \cdot F_{j\textbf{U}}$ is symmetric if and only if there is a smooth function $J^f_j(\textbf{U})$  such that $s_\textbf{U}F_{j\textbf{U}}=J^f_{j\textbf{U}}$. Thus with the help of condition (\ref{condition 1}), the entropy flux can be rewritten into a full divergent form as we have seen in \eqref{ent-diss}.

In non-equilibrium systems, $\textbf{q}{(\textbf{U})}$ represents the source or sink terms due to chemical reactions, radiation, electric dissipation and \textit{etc.}
How to properly incorporate them into the balance equations is regarded as a challenging and extraordinary difficult problem \cite{jozsa2020general}.
Here the second condition (\ref{condition 2}) of CDF provides an elegant solution to this problem based on a generalized nonlinear version of Onsager's relation between thermodynamic fluxes and forces \cite{Peng2018Generalized,de2013}, since $\textbf{M}(\textbf{y},\textbf{z})$ is no longer a constant matrix and can also depend on dissipative variables. Furthermore, in the classical Onsager's relation, $\textbf{M}$ has to be symmetric in order to keep the time reflection symmetry. While in our case, this restriction is abandoned too in order to incorporate more complicated situations, like open systems, non detailed balance conditions.

The last condition  (\ref{condition 3}) is usually referred to as the Kawashima condition in literature, which is satisfied by many classical hyperbolic-parabolic type systems, including the Navier-Stokes equations, many discrete velocity models of the Boltzmann equation, \textit{etc.} \cite{shizuta1985systems}

With respect to above conditions, a global existence theorem \cite{Yong2004entropy} on the unique solution to Eq.  \eqref{con-diss} could be rigorously established, providing the initial data close to their equilibrium values. This global existence theorem not only provides the mathematical foundation for our CDF, but also makes a major difference from other physical based modeling approaches. If relaxing the requirement on solutions from global existence to local existence, the Kawashima condition is not needed any more. Actually, unlike the entropy dissipation conditions in  (\ref{condition 1}) and (\ref{condition 2}), contrary examples unsatisfying the Kawashima condition but still enjoying global smooth solutions have been reported, \textit{e.g.} the gas dynamics in thermal non-equilibrium \cite{zeng1999gas}. Thus how to weaken the condition (\ref{condition 3}) and find alternative new general conditions would be of great interest.

Rooted also in the Godunov structure for hyperbolic systems, the Symmetric Hyperbolic Thermodynamically Compatible Framework \cite{peshkov2018continuum} (SHTC) is another way to model irreversible processes, whose mathematical rigorousness on the local well-posedness of the Cauchy problem has been clarified. However, in contrast to CDF, SHTC does not put any restriction on the source term $\textbf{Q}(\textbf{U})$. The missing of structural stability conditions in (\ref{condition 2})-(\ref{condition 3}) may give rise to instability in hyperbolic systems, such as the BISQ models discussed in Sec. \ref{BISQ} below, which therefore does not satisfy the general requirement of thermodynamics.

\subsection{Gradient flows in the absence of source terms}
The structural conditions of CDF presented in the last section highlight an intrinsic coupling between the flux terms and source terms, which guarantees the existence of global smooth solutions. However, in the absence of source terms, \textit{i.e.}, $\textbf{Q}(\textbf{U})=0$, the CDF structure in \eqref{con-diss} reduces to much simpler local conservation laws studied a lot in continuum mechanics and hydrodynamics.

In particular, we focus on the case when the probability density function is taken as the only state variable, $\textbf{U}=(\rho(\textbf{x},t))$, which is assumed to be absolutely continuous. Jordan, Kinderlehrer and Otto made an important discovery that any smooth positive solution of
\begin{equation}\label{gradient}
  \partial_t \rho+\nabla\cdot \bigg(\rho\nabla\frac{\delta F}{\delta \rho}\bigg)=0
\end{equation}
can be seen as a trajectory of the gradient flow associated with the free energy $F(\rho)$ in the Wasserstein space \cite{Villani2003,Villani2014}.

Based on different physical meanings, the free energy can be separated into three basic kinds of energies: the internal energy $\textrm{U}(\rho)=\int_{{\mathbb{R}^\Lambda}}U(\rho(\textbf{x},t))d\textbf{x}$, the potential energy $\textrm{V}(\rho)=\int_{\mathbb{R}^\Lambda}\rho(\textbf{x},t)V(\rho(\textbf{x},t))d\textbf{x}$ and the interaction energy $\textrm{W}(\rho)=\frac{1}{2}\int_{\mathbb{R}^\Lambda \times \mathbb{R}^\Lambda}W(\textbf{x}-\textbf{y})d\rho(\textbf{x})d\rho(\textbf{y})$. If $F=\textrm{U}+\textrm{V}+\textrm{W}$, the corresponding gradient flow is
\begin{equation}
  \partial_t \rho=\nabla\cdot \bigg[\rho\nabla U'(\rho)+\rho\nabla V+\rho(\rho\ast\nabla W)\bigg].
\end{equation}
An incomplete list of partial differential equations which fall into above gradient flow systems include: the heat equation with $U=\rho\ln\rho, V=W=0$, a porous medium type equation with $U=\rho^m/(m-1), V=W=0$, the linear Fokker-Planck equation with $U=\rho\ln\rho, V=V(\textbf{x}), W=0$, \textit{etc.}

A most significant mathematical property of the gradient flow system in \eqref{gradient} is the Lyapunov functional nature of $F(\rho)$. Whenever $\rho(\textbf{x},t)$ is a solution, $F(\rho(\textbf{x},t))$ is a nonincreasing function of time $t$, since it is straightforward to show that
\begin{equation}
-\frac{d}{dt}F(\rho)=\int_{{\mathbb{R}^\Lambda}}\rho\bigg|\nabla\frac{\delta F}{\delta\rho}\bigg|^2d\textbf{x},
\end{equation}
which is known as the dissipation rate of free energy (or entropy production rate).

Now a key problem interested in both non-equilibrium thermodynamics and mathematics is the trend to equilibrium in an entropy sense, meaning
\begin{equation}
F(\rho(\textbf{x},t))\xrightarrow[t\rightarrow\infty]{} F(\rho(\textbf{x},\infty))
\end{equation}
In many cases, the convergence in an entropy sense implies convergence of the solution in the $L^1$ norm. For example, we have $F(\rho(\textbf{x},t))-F(\rho(\textbf{x},\infty))\geq\frac{1}{2}\|\rho(\textbf{x},t)-\rho(\textbf{x},\infty)\|^2_{L^1}$ for the Fokker-Planck equation by Csiszar-Kullback-Pinsker inequality. The proof of the entropy convergence is generally related to an entropy-entropy production inequality \cite{Villani2003,Villani2014},
\begin{equation}
-\frac{d}{dt}F(\rho)\geq\Phi[F(\rho(\textbf{x},t)-F(\rho(\textbf{x},\infty)],
\end{equation}
where $\Phi$ is continuous and strictly increasing from $0$. Furthermore the form of $\Phi({x})$ is closely related to the convergence rate. For example, if $\Phi({x})=k{x}$, the entropy will approach its limit value exponentially fast; contrarily if $\Phi({x})={x}^{\alpha}$ with $\alpha>1$, the convergence rate will be algebraic.

Based on this routine, the convergence problems of many gradient flow systems \cite{Villani2014,Santambrogio2015}, like the Fokker-Planck equation, porous media equations with drift, Vlasov equations and so on, were investigated systematically and became one of leading research fields in mathematical physics and partial differential equations.

\subsection{A typical example: the generalized Newton-Stokes-Fourier's law}\label{non-NSF}
Before immersing into applications in various fields, we first present a typical example -- the generalized Newton-Stokes-Fourier's law step by step, to provide the readers a complete and detailed view on how constitutive relations are derived from CDF, and what kinds of advantages are reflected during the derivation.

Let's consider a one-component  system of compressible fluids in non-isothermal environments. The conservation laws of mass, momentum and total energy read respectively
\begin{align}
&\frac{\partial}{\partial t}\rho + \nabla \cdot (\rho \textbf{v})=0,\label{mass}\\
&\frac{\partial}{\partial t}(\rho \textbf{v})+ \nabla\cdot(\rho \textbf{v} \otimes \textbf{v})=\nabla \cdot \bm{\sigma},\label{linmom}\\
&\frac{\partial}{\partial t}(\rho e) + \nabla \cdot(\rho \textbf{v} e)=\nabla \cdot (\bm{\sigma} \cdot \textbf{v} - \textbf{q})\label{energy},
\end{align}
where $(\textbf{x},t) \in \mathbb{R}^{\Lambda} \times \mathbb{R}^1_{\geq 0}$, the spatial gradient operator $\nabla=(\partial x_1, \partial x_2, \cdots, \partial x_{\Lambda})^T$. $\otimes$ denotes the tensor product. $\rho, u, e=u+|\textbf{v}|^2/2$ represent the fluid density, momentum and specific energy separately. $\bm{\sigma}$ is the stress tensor, and $\textbf{q}$ is the heat flux.

To find constitutive relations for the stress tensor $\bm{\sigma}$ and heat flux $\textbf{q}$, and thus close the PDE system given above, CIT adopts the Newton's law for viscosity and Fourier's law for heat conduction. The conservation laws together with the classical constitutive relations are known as the Navier-Stokes-Fourier (NSF) equations in classical hydrodynamics. However, the NSF equations fail to describe many interesting phenomena in viscoelastic fluids. So based on CDF, a generalized hydrodynamic system was deduced by Zhu \textit{et al.} \cite{zhu2015conservation}.
Recall that, in EIT the state variable space for this system is directly enlarged to be a combination of $(\nu, u)$ and $(\textbf{P}, \textbf{q})$, with the specific volume $\nu=1/\rho$ and internal energy $u$. However, CDF adopts the unspecified pairs $(\textbf{C},\textbf{w})$ in replace of $(\textbf{P}, \textbf{q})$ to be dissipative variables, where $(\textbf{C}, \textbf{w})$ have the same sizes as $(\textbf{P}, \textbf{q})$.
That is, $\textbf{C}$ is a $\Lambda\times \Lambda$ matrix, and vector $\textbf{w}\in \mathbb{R}^{\Lambda}$.
Assume there is a strictly concave function w.r.t. $(\nu,u,\textbf{w},\textbf{C})$,
\begin{equation}\label{entropy0}
s=s(\nu, u, \textbf{w},\textbf{C}),
\end{equation}
which is also known as the non-equilibrium specific entropy.

It is noticeable that, the generalized non-equilibrium entropy in Eq. \eqref{entropy0} would reduce to the classical entropy in equilibrium, $s|_{eq}=s_0(\nu, u)$.
Consequently, the equilibrium temperature is defined as a partial derivative of the equilibrium entropy, $T^{-1}=\frac{\partial s_0}{\partial u}$.
The concept of (equilibrium) temperature is a direct consequence of the zeroth law of thermodynamics. While its absolute value, independent of the material properties of the system under study, is offered by the first and second laws of thermodynamics, like in the Carnot's cycle.
For this non-isothermal system, the non-equilibrium temperature $\theta$ and non-equilibrium thermodynamic pressure $p$ are introduced self-consistently as
\begin{equation}\label{non-eq temp}
\theta^{-1}=\frac{\partial}{\partial u} s(\nu, u, \textbf{w},\textbf{C}), \quad
\theta^{-1}p=\frac{\partial}{\partial \nu} s(\nu, u, \textbf{w},\textbf{C}).
\end{equation}
\begin{remark}
The definition of equilibrium temperature $T$, is borrowed by CIT and is directly applied to non-equilibrium conditions based on the hypothesis of local equilibrium.
In contrast, definitions and measurements of temperature out of the local equilibrium involve many subtle and non-trivial conceptual issues, and are still open.
Several theoretical models have been designed to modify or extrapolate the definition of equilibrium temperature, including the generalized non-equilibrium temperature in EIT \cite{jou1996extended}, the contact temperature based on axiomatic bases \cite{Muschik1977Empirical}, the ones in kinetic theory, information theory and stochastic processes.
The non-equilibrium temperature $\theta$ of CDF in \eqref{non-eq temp} is analogous to that in EIT, except for the arguments of entropy.
Interested readers are referred to the review by Casas-V{\'a}zquez and Jou \cite{Casas2003Temperature} and references therein, for the conceptual changes of non-equilibrium temperature, their practical applications and possible experiments.
\end{remark}

Now the time evolution of the entropy density (per unit volume) reads
\begin{align*}
&{\partial_t}(\rho s) + \nabla \cdot (\rho \textbf{v} s) \\
=&\theta^{-1} (p \nabla \cdot \textbf{v} - \nabla \cdot \textbf{q} + \bm{\sigma}^T : \nabla \textbf{v})\\
&
+ s_{\textbf{w}} \cdot [ {\partial_t}(\rho \textbf{w}) + \nabla \cdot (\rho \textbf{v} \textbf{w}) ]
+ s_{\textbf{C}}^T:    [ {\partial_t}(\rho \textbf{C}) + \nabla \cdot (\rho \textbf{v} \textbf{C}) ] \\
=&-\nabla \cdot (\theta^{-1}\textbf{q}) + \textbf{q} \cdot \nabla {\theta}^{-1} + {\theta}^{-1}\bm{\tau}^{T}:\nabla \textbf{v} \\
&
+ s_{\textbf{w}} \cdot [ {\partial_t}(\rho \textbf{w}) + \nabla \cdot (\rho \textbf{v} \textbf{w}) ]
+ s_{\textbf{C}}^T:    [ {\partial_t}(\rho \textbf{C}) + \nabla \cdot (\rho \textbf{v} \textbf{C}) ] \\
=&-\nabla \cdot (\theta^{-1}\textbf{q}) \\
&
+ \textbf{q} \cdot [ {\partial_t}(\rho \textbf{w}) + \nabla \cdot (\rho \textbf{v} \textbf{w}) + \nabla \theta^{-1}]
+ {\theta}^{-1}\bm{\tau}^{T}: [ {\partial_t}(\rho \textbf{C}) + \nabla \cdot (\rho \textbf{v} \textbf{C}) + \nabla \textbf{v}] \\
=&-\nabla \cdot \textbf{J}^f + \Sigma^f,
\end{align*}
where we denote $\bm{\tau}=\bm{\sigma}+p\textbf{I}$ in the second step, and use the relations $s_{\textbf{w}}=\textbf{q}$ and $s_{\textbf{C}}=\theta^{-1}\bm{\tau}$ in the third step. In the last line, $\textbf{J}^f=\theta^{-1}\textbf{q}$ and
\begin{equation}
\Sigma^f=
\textbf{q} \cdot [ {\partial_t}(\rho \textbf{w}) + \nabla \cdot (\rho \textbf{v} \textbf{w}) + \nabla \theta^{-1}]
+ {\theta}^{-1}\bm{\tau}^{T}: [ {\partial_t}(\rho \textbf{C}) + \nabla \cdot (\rho \textbf{v} \textbf{C}) + \nabla \textbf{v}]
\end{equation}
represent the entropy flux and entropy production rate.
It is noticeable that, the adoption of conjugate variables offers us more freedom to deduce genuinely nonlinear constitutive relations.
Assume the stress tensor $\bm{\tau}$ (and consequently $\textbf{C}$) to be symmetric, we have
\begin{equation}\label{constitutive of general type}
\begin{pmatrix}
& {\partial_t}(\rho \textbf{w}) + \nabla \cdot (\rho \textbf{v} \textbf{w}) + \nabla \theta^{-1}  \\
& {\partial_t}(\rho \textbf{C}) + \nabla \cdot (\rho \textbf{v} \textbf{C}) + \textbf{A}
\end{pmatrix}
=\textbf{M} \cdot
\begin{pmatrix}
& \textbf{q} \\
& \theta^{-1} \bm{\tau}
\end{pmatrix}
,
\end{equation}
based on CDF. Here the dissipation matrix $\textbf{M}(\nu, u, \textbf{w}, \textbf{C})$ is positive semi-definite to guarantee the non-negativity of the entropy production rate $\Sigma^f \geq 0$, and $\textbf{A}=(\nabla \textbf{v} + \nabla \textbf{v}^T)/2$.
Now the generalized hydrodynamic equations for non-Newtonian fluids are completed, which was first presented in Ref. \cite{zhu2015conservation}.

Recall the non-equilibrium entropy function in Eq. \eqref{entropy0} would reduce to the classical entropy in equilibrium, $s|_{\textbf{w}=\textbf{0},\textbf{C}=\textbf{0}}=s_0(\nu, u)$. To get a concrete idea on the constitutive relations \eqref{constitutive of general type}, Zhu \textit{et al.} \cite{zhu2015conservation} specified the entropy function and dissipation matrix as
\begin{equation}
s=s_0(\nu, u) - \frac{1}{2\nu\epsilon_0}|{\textbf{w}}|^2 - \frac{1}{2\nu\epsilon_1}|\dot{{\textbf{C}}}|^2
 - \frac{1}{2\nu\epsilon_2}|{\mathring{\textbf{C}}}|^2,
\end{equation}
and
\begin{equation}
\textbf{M}\cdot  \left(\begin{array}{c}
    {\textbf{q}}\\[3mm]
    \theta^{-1}\bm{\tau}
  \end{array}\right)= \left(\begin{array}{c}
\frac{1}{\theta^2\lambda}{\textbf{q}}\\[3mm]
\frac{\dot{\bm{\tau}}}{\xi}
+\frac{\mathring{\bm{\tau}}}{\kappa}
\end{array}\right),
\end{equation}
where $\dot{\textbf{C}}=\frac{1}{\Lambda}\hbox{tr}(\textbf{C})\textbf{I}$, $\mathring{{\textbf{C}}}=\frac{1}{2}(\textbf{C}+\textbf{C}^T)-\frac{1}{\Lambda}\hbox{tr}(\textbf{C})\textbf{I}$. Parameters $\epsilon_i>0$ are related to the different relaxation times $(i=0,1,2)$, $\lambda>0$ is the heat conduction coefficient, and $\xi>0, \kappa>0$ are viscosity parameters. With above choices, we have a concrete form of Eq. \eqref{constitutive of general type}
\begin{equation}\label{constitutive of concrete type}
\left\{
             \begin{array}{lr}
             \epsilon_0\big[\partial_t \textbf{q} +\nabla\cdot(\textbf{v}\textbf{q})\big] -\nabla\theta^{-1}=-\frac{\textbf{q}}{\theta^2\lambda},&\\
             \epsilon_1\big[\partial_t(\theta^{-1}\dot{\bm{\tau}})+
             \nabla\cdot(\theta^{-1}\textbf{v}\dot{\bm{\tau}})\big]
             -\dot{\nabla\textbf{v}}=-\frac{\dot{\bm{\tau}}}{\kappa},&\\
             \epsilon_2\big[\partial_t(\theta^{-1}\mathring{\bm{\tau}})+
             \nabla\cdot(\theta^{-1}\textbf{v}\mathring{\bm{\tau}})\big]
             -\mathring{\nabla\textbf{v}}=-\frac{\mathring{\bm{\tau}}}{\xi}, &
             \end{array}
\right.
\end{equation}
which are the Cattaneo's law for heat conduction and Maxwell's laws for viscoelasticity. Formally, these relaxation-type constitutive relations reduce to the classical NSF relations,
$$
\textbf{q}=-\lambda\nabla \theta, \quad
\bm{\tau}=  \xi\mathring{\nabla\textbf{v}} + \kappa\dot{\nabla\textbf{v}},
$$
through Maxwellian iteration, as $\epsilon$'s go to zero, .

Now a natural question arises. Is the CDF structure for non-Newtonian fluids compatible with the classical NSF equations in the relaxation limit? Mathematically, this problem is highly non-trivial.
As an important advantage of CDF, the compatibility between our generalized hydrodynamics and classical NSF equations can be rigorously proved, provided two natural assumptions:
\begin{itemize}
 \item \textbf{Compatibility Assumption}: at local equilibrium, \textit{i.e.}, $s_{\textbf{w}}=0$ and $s_{\textbf{C}}=0$, we have
$$ s(\nu, u, \textbf{w}, \textbf{C})=s_{0}(\nu, u), \quad \textbf{M}(\nu,u,\textbf{w}, \textbf{C})=(\textbf{K}^{FNS})^{-1}.$$
Here $\textbf{K}^{FNS}$ denotes the dissipation matrix at local equilibrium, such that
\begin{equation*}
\textbf{K}^{FNS} \cdot
\begin{pmatrix}
\nabla T^{-1}\\
-\frac{1}{2}(\nabla \textbf{v} +\nabla \textbf{v}^T)
\end{pmatrix}
=\begin{pmatrix}
-\lambda \nabla T\\
-T^{-1}
\left[
\xi\mathring{(\nabla \textbf{v})}+ \kappa (\dot{ \nabla \textbf{v} })
\right]
\end{pmatrix}
.
\end{equation*}
 \item \textbf{Causality Assumption}: let $\textbf{z}=\{\textbf{w}, \textbf{C}\}$, then
$$
\textbf{z}=\textbf{0}, \quad \hbox{if} \quad s_\textbf{z}=\textbf{0}.
$$
\end{itemize}

The compatibility condition is a natural prerequisition to keep the consistency of the two systems at local equilibrium, while the casuality condition ensures that for any given thermodynamic conjugate pair $(\textbf{z}, s_{\textbf{z}})$, if one is zero then the other must be zero as well. As the relaxation time goes to zero, smooth solutions to the generalized hydrodynamic equations
converge to that of the NSF equations in a proper Sobolev space, as stated through the following theorem.

\begin{theorem}
(Compatibility of the Generalized and Classical Hydrodynamics)
Under the compatibility and casuality assumptions, suppose the density $\rho$, velocity $\textbf{v}$ and energy $e$ of the classical hydrodynamic system are continuous and bounded in $(\textbf{x},t)\in\Omega\times[0,t_*]$ with $t_*<\infty$, and further satisfy $\displaystyle{\inf_{\textbf{x},t}\rho(\textbf{x},t)}>0$ and
$$
\rho, \textbf{v}, e \in C([0,t_*],H^{s+3})\cap C'([0,t_*],H^{s+1}(\Omega)),
$$
with integer $s\geq [{\Lambda}/2]+2$. Then there exist positive numbers $\epsilon_0=\epsilon_0(t_*)$ and $K=K(t_*)$ such that for $\epsilon\leq \epsilon_0$ the generalized hydrodynamic system, with initial data
in $H^s(\Omega)$ satisfying $\|(\rho^\epsilon,\rho^\epsilon \textbf{v}^\epsilon,\rho^\epsilon e^\epsilon)|_{t=0}-(\rho,\rho \textbf{v},\rho e)|_{t=0}\|_s=O(\epsilon^2)$, has a unique classical solution satisfying
$$(\rho^\epsilon,\rho^\epsilon \textbf{v}^\epsilon,\rho^\epsilon e^\epsilon,\rho^\epsilon  \textbf{w}^\epsilon,\rho^\epsilon \textbf{C}^\epsilon)\in C([0,t_*],H^s(\Omega))$$
and
\begin{equation*}
\mathop {\sup }\limits_{t \in [0,{t_*}]}\|(\rho^\epsilon,\rho^\epsilon \textbf{v}^\epsilon,\rho^\epsilon e^\epsilon)-(\rho,\rho \textbf{v},\rho e)\|_s\leq K(t_*)\epsilon^2.
\end{equation*}
\end{theorem}

\section{Classical Models in Mathematical Physics}
\label{applications I}
It is observed that many classical models in mathematical physics can be reformulated into the elegant form of CDF. This observation not only justifies the fact that CDF indeed grasps the common mathematical structure behind diverse models, but also serves as a cornerstone for the application of CDF to various irreversible processes.

In this section, we are going to review the conservation-dissipation structure of master equations, Fokker-Planck (F-P) equations, chemical mass-action equations, moment hierarchies of Boltzmann equations and many other classical models in mathematical physics. Based on the intrinsic connections among these model equations, we will present our results following three routines as:

Routine 1 (stochastic models): \textbf{Master equations $\rightarrow$ Fokker-Planck (F-P) equations $\rightarrow$ Chemical mass-action euqations};

Routine 2 (hydrodynamic systems): \textbf{Moment hierarchies of Boltzmann equation $\rightarrow$ Euler equations
$\rightarrow$ Navier-Stokes-Fourier (NSF) equations $\rightarrow$ Non-NSF equations};

Routine 3 (optics, radiation and \textit{etc.}): \textbf{Quasi-linear Maxwell's equations for nonlinear optics $\rightarrow$ Radiation hydrodynamics $\rightarrow$ Chemically reactive flows}.


\subsection{Stochastic models} \label{example1}
In the first case, we focus on a discrete Markov process with finite states characterised by general master equations in the form of
\begin{eqnarray}\label{mastereq1}
\frac{d}{dt}p_i(t)=\sum_{j\neq i}(q_{ij}p_j-q_{ji}p_i),\quad i=1,\cdots,N,
\end{eqnarray}
where $p_i\geq0$ indicates the probability for finding the system at state $i$, and $q_{ij}\geq0$ $(i\neq j)$ denotes the transition rate from state $j$ to $i$. In particular, $q_{ii}=-\sum_{j\neq i}q_{ji}$. 
Denote the steady state of master equations \eqref{mastereq1} as $\{p_i^s, 0<p_i^s<1 \}$ such that $\sum_{j} q_{ij}p^s_j  =\sum_{j} q_{ji}p^s_i,~ i=1,\cdots,N$.
If $(q_{ij})$ further satisfies the condition of detailed balance, \textit{i.e.}, $q_{ij}p^e_j  =q_{ji}p^e_i$, the steady state is at equilibrium. The master equations have been extensively applied to quantum thermodynamics \cite{Vinjanampathy2016quantum}, open chemical reactions \cite{ge2016mesoscopic,Rao2016Nonequilibrium}, molecular biology \cite{van1983}, an so on. In Ref. \cite{Peng2018Generalized} it was found that master equations \eqref{mastereq1} has a structure of CDF, which is stated as follows.
\begin{theorem}
(CDF for Master Equations) Given the thermodynamic flux $J_i=\frac{dp_i}{dt}$ and force $X_i=-\frac{\partial F}{\partial p_i}$, where $F=\sum_ip_i\ln(p_i/p_i^s)$, the master equation \eqref{mastereq1} obeys the CDF:
\begin{eqnarray}\label{M_master}
J_i=\sum_jM_{ij}X_j,
\quad
M_{ij}= \left\{ \begin{array}{ll}
-\frac{q_{ij}\exp[\sigma(p_i,p_j)]}{k_BTp_i^s},\qquad\qquad i\neq j,\\
\sum_{j\neq i}\frac{q_{ij}\exp[\sigma(p_i,p_j)]}{k_BTp_i^s},\qquad~ i=j.
\end{array} \right.
\end{eqnarray}
$M_{ij}$ in \eqref{M_master} is a positively stable matrix. Zero is single eigenvalue of $\textbf{M}$ and its right null space is spanned by $(1/\sqrt{N},\cdots,1/\sqrt{N})$, which is independent of $p_i$ ($i=1,2,\cdots,N$). Furthermore, $\textbf{M}$ is symmetric and positive semi-definite if and only if (iff for short) the condition of detailed balance holds.
\end{theorem}

%
Please see Ref. \cite{Peng2018Generalized} for a proof. It is noticeable that $\textbf{M}$ does not need to be symmetric. The anti-symmetric part $\textbf{M}^a=(\textbf{M}-\textbf{M}^T)/2$ actually plays a key role in measuring how far a system is kept away from the equilibrium state. Therefore, CDF is closely related with the steady state thermodynamics of master equations, which recently becomes a rapidly growing field \cite{ge2010physical,esposito2010}.

In master equations, transitions among all possible states are involved, which leads to a heavy burden in both modeling and computing.
The F-P equation makes a simplification by expanding the transition rate matrix and only keeping the first two leading moments \cite{van1983}. We have shown that the F-P equation
\begin{equation}\label{fokker}
\frac{\partial p}{\partial t}=-\sum_{i=1}^{\Lambda} \frac{\partial}{\partial x_i} {K_i}(\textbf{x},t), \quad
K_i=u_i(\textbf{x})p-\sum_{j=1}^{\Lambda}D_{ij}(\textbf{x})\frac{\partial p}{\partial x_j}
\end{equation}
can also be casted into CDF \cite{Peng2018Generalized}, where the vector $\textbf{u}=(u_1,u_2,\cdots,u_{\Lambda})$ and the symmetric and positive-definite matrix $\textbf{D}=(D_{ij})_{{\Lambda} \times {\Lambda}}$ denote the respective drift and diffusion coefficients \cite{risken1996fokker}.
Referring to the probability density of steady state $p^s(\textbf{x})$ and adopting the Tsallis relative entropy with non-extensive parameter $\alpha=2$
\begin{equation}
F_2(t)= \frac{1}{2}\int d\textbf{x} \frac{p(\textbf{x},t)^2}{p^s(\textbf{x})}-\frac{1}{2},
\end{equation}
we have the following results.
\begin{theorem} (CDF for F-P Equations)
Given the thermodynamic flux $J(\textbf{x},t)=\frac{\partial p}{\partial t}$ and the thermodynamic force $X(\textbf{x},t)=-\frac{\delta F_2}{\delta p(\textbf{x},t)}$, the F-P equation \eqref{fokker} obeys the CDF:
\begin{equation}
J(\textbf{x},t)=\int d\textbf{y}  M(\textbf{x},\textbf{y}) X(\textbf{y},t),
\end{equation}
with the kernel $M(\textbf{x},\textbf{y})$ defining as
\begin{eqnarray}
{M}(\textbf{x},\textbf{y})=\nabla_{\textbf{x}} \cdot \left[\bm{D}(\textbf{x})p^s(\textbf{x}) \cdot \nabla_{\textbf{y}} \delta(\textbf{y}-\textbf{x})\right]- \bm{K}^s(\textbf{x}) \cdot \nabla_\textbf{y} \delta(\textbf{y}-\textbf{x}),
\end{eqnarray}
where $\delta(\textbf{y}-\textbf{x})$ is the Dirac's delta function and
$\nabla_\textbf{y} \equiv (\frac{\partial }{\partial y_1}, \frac{\partial }{\partial y_2}, \cdots, \frac{\partial }{\partial y_{\Lambda}})$.
The integral operator defined in space $L_2(\mathbb{R}^{\Lambda})$ with kernel $M(\textbf{x},\textbf{y})$ being positively stable. If the F-P equation further satisfies the detailed balance condition, $M(\textbf{x},\textbf{y})$ becomes symmetric and positive definite.
\end{theorem}

Here we adopt the Tsallis entropy to derive CDF of F-P equations, however, this intrinsic structure is not restricted to the non-extensive entropy. For example, Dong \cite{Dong2016Conservation} explored the F-P equations based on the Boltzmann-Gibbs entropy by studying the underlying ordinary differential equations in the phase space.

In stochastic chemical reactions, the transition rates are specified through propensity functions, which results in the chemical master equations (CME).
And it is well-known that the expectation of number density in CME leads to the concentration in chemical mass-action equations, which
was first proved by Kurtz in the limit of large system size \cite{kurtz1972}.

Without loss of generality, we consider a chemical reaction system with $N$ species and $M$ reversible reactions
\begin{equation}\label{reaction}
\ce{\nu_{i1}^+ S_1 + \nu_{i2}^+ S_2 + \cdots + \nu_{iN}^+ S_N
 <=>T[$\kappa_i^+$][$\kappa_i^-$] \nu_{i1}^- S_1 + \nu_{i2}^- S_2 + \cdots + \nu_{iN}^- S_N},
\end{equation}
for $i=1,2,\cdots,M$. According to laws of mass-action, the concentration of k'th species evolves as
\begin{equation}\label{mass-action equation}
\frac{dc_k}{dt}=\sum_{i=1}^{M}(\nu_{ik}^+ - \nu_{ik}^-)\left(\kappa_{i}^- \prod_{j=1}^N {c_j}^{\nu_{ij}^-} - \kappa_i^+ \prod_{j=1}^N {c_j}^{\nu_{ij}^+} \right),
\end{equation}
for $k=1,2,\cdots, N$, where $c_k(t)=[S_k]$ is the concentration for the $k$th species, $\nu_{ik}^+$ and $\nu_{ik}^-$ are stoichiometric coefficients for the $k$th species in the $i$th reaction, $\kappa_i^+$ and $\kappa_i^-$ are the forward and backward reaction rate constants, respectively.
Under the condition of detailed balance, Yong presented the CDF of the mass-action equations \cite{yong2012conservation}. In this case, the free energy is chosen as $F(t)=\sum_{i=1}^N \left( c_i \ln c_i - c_i \ln c_i^e -c_i + c_i^e \right)$, and $\{c_i^e\}$ is the equilibrium state satisfying the condition of detailed balance.

\begin{theorem}
(CDF for Mass-Action Equations) Given the thermodynamic flux $J_i=\frac{dc_i}{dt}$ and force $X_i=-\frac{\partial F}{\partial c_i}$, the mass-action equation \eqref{mass-action equation} obeys the CDF:
\begin{eqnarray}\label{Mass_master}
J_i=\sum_jM_{ij}X_j,
\end{eqnarray}
where $\textbf{M}(\textbf{c})=(M_{ij})$ is a symmetric and positive semi-definite matrix, and the null space of $\textbf{M}$ is independent of $\{c_i >0\}_{i=1}^{N}$.
\end{theorem}

Notice that the positively stable matrix of master equations in \eqref{M_master} becomes symmetric and positive semi-definite iff the detailed balance condition is satisfied, which is exactly the case studied in chemical mass-action equations. How to generalize the result to the complex balance condition, as that for master equations, is still open.

Based on the CDF of mass-action equations, Yong further justified the mathematical validity of partial equilibrium approximation (PEA) \cite{yong2012conservation}, which is frequently used in model reduction.
By separating reactions into fast and slow ones, and denoting a small parameter $\varepsilon~(0<\varepsilon\ll 1)$ to measure the fastness, we can rewrite the chemical mass-action equations as
\begin{equation}\label{mass-action equation2}
\frac{dc_k}{dt}=-\frac{1}{\varepsilon} \sum_jM_{kj}\frac{\partial F}{\partial c_j} + p_k(\textbf{c}), \quad k=1,2,\cdots, N,
\end{equation}
where $p_k(\textbf{c})$ denotes the slow reactions.
Then according to the singular perturbation theory, the validity of PEA method can be rigorously justified as follows \cite{yong2012conservation}.
\begin{theorem}
(Validity of PEA) The solution to an initial-value problem of the two-scale system \eqref{mass-action equation2} uniformly converges, in any bounded-time interval away
from $t=0$,  to the algebraic equation $\sum_{i=1}^{M}(\nu_{ik}^+ - \nu_{ik}^-)\left(\kappa_{i}^+ \prod_{j=1}^M {c_j}^{\nu_{ij}^-} - \kappa_i^- \prod_{j=1}^M {c_j}^{\nu_{ij}^+} \right)=0$ and the corresponding simplified system, as $\varepsilon \rightarrow 0$.
\end{theorem}


Considering the mathematical connections among the master equations, F-P equations and mass-action equations, it has been further justified that the thermodynamic formalism constructed above these equations are also consistent \cite{peng2018markov, ge2016nonequilibrium}. Moreover, all these equations can be cast into the CDF, which reflects an intrinsic dissipative structure among them.

\subsection{Hydrodynamic systems} \label{example1}
In the previous section, we have shown that the generalized Navier-Stokes-Fourier equations could be derived from CDF. Actually, many other classical hydrodynamic systems also fall into the category of CDF. This fact could be more clearly learnt from the moment hierarchies of Boltzmann equation, a systematic and rigorous way to derive hydrodynamic equations from mesoscopic kinetic theories, like the famous Chapman-Enskog expansion \textit{etc. }

Hong \textit{et al.} \cite{hong2015novel} studied the Levermore's moment hierarchies of Boltzmann equation within the framework of CDF, which reads
\begin{equation}\label{boltzmann}
\frac{\partial}{\partial t} \int c_i f d \bm{\xi} + \nabla \cdot \int \bm{\xi} c_i f d\bm{\xi}
=\int c_i (f_{\star} f_{\star}' - f f') B(\bm{w},\bm{\xi}, \bm{\xi'} )d\bm{w}d\bm{\xi'}d\bm{\xi},
\end{equation}
for each $i=0,1,\cdots, n-1$, where $\int c_i f d \bm{\xi}$ is the $i$-th moment with $\{c_i=c_i(\bm{\xi}), i=0,1,\cdots, n-1\}$ being an admissible space. Here $\bm{\xi}$ and $\bm{\xi}'$ denote the particle velocities before binary collisions, while $\bm{\xi}_{\star}$ and $\bm{\xi}_{\star}'$ after the collision. The respective distribution functions of particles are $f=f(\bm{r}, \bm{\xi},t)$, $f=f(\bm{r}, \bm{\xi}',t)$, $f_{\star}=f_{\star}(\bm{r}, \bm{\xi}_{\star},t)$ and $f_{\star}'=f_{\star}'(\bm{r}, \bm{\xi}_{\star}',t)$, the scattering angle is $\bm{w}$, the collision kernel $B(\bm{w}, \bm{\xi}, \bm{\xi}')=B(\bm{w},\bm{\xi}',\bm{\xi})=B(\bm{w}, \bm{\xi}_{\star}, \bm{\xi}'_{\star})$ is positive almost everywhere in its domain.

Choose the moments $\phi_i=\int c_i f d\bm{\xi}~(i=0,1,\cdots, n-1)$ in Eq. \eqref{boltzmann} as state variables, and specify the Boltzmann entropy $s=-k_B \int (f\ln f - f) d\bm{\xi}$.
Here the ansatz \cite{Levermore1996Moment}
\begin{equation*}
f=f(\bm{r}, \bm{\xi},t)=\exp\bigg[\sum_{i=0}^{n-1} c_i(\bm{\xi}) \odot \alpha_i\bigg],
\end{equation*}
is adopted with $\alpha_i$ denoting the $i$-th tensors that can be determined via normalization conditions. It was shown that the Levermore's moment-closure hierarchies for the Boltzmann equation possessed the CDF in \eqref{con-diss} as
\begin{align}\label{hir}
\textbf{U}=\begin{pmatrix} \phi_0\\ \phi_1 \\ \vdots \\ \phi_{n-1} \end{pmatrix}, \quad \textbf{F}(\textbf{U})=\begin{pmatrix} \phi_1\\ \phi_2 \\ \vdots \\ \phi_n \end{pmatrix}, \quad \textbf{Q}(\textbf{U})=\textbf{M}(\textbf{U}) \cdot s_\textbf{U},
\end{align}
where $\textbf{M}(\textbf{U})$ is symmetric, positive semi-definite, and its null space is independent of the state variable $\textbf{U}$.

Now we are at a position to derive macroscopic hydrodynamic equations from the mesoscopic Boltzmann equation.
As a straightforward application, a new ten-moment model beyond the classical hierarchies
is derived, which recovers the Euler equations in the equilibrium state.
We take the moments up to order 2, \textit{i.e.}, the density $\phi_0=\int f d\bm{\xi}\equiv \rho$, the momentum $\phi_1=\int \bm{\xi}f d\bm{\xi}\equiv \rho \textbf{v}$, and the second-order stress tensor $\phi_2=\int \bm{\xi} \otimes \bm{\xi} f d\bm{\xi} \equiv \textbf{P}$.
The entropy is specified as $s=k_B \rho [\ln \det \bm{\Theta} - 2\ln \rho + 3 \ln (2\pi) +3]/2$, with $\bm{\Theta}=\rho^{-1}\textbf{P} - \textbf{v}\otimes \textbf{v}$.
It can be proved that the moment-closure equations \eqref{hir} become
\begin{align*}
&\frac{\partial}{\partial t}\rho + \nabla \cdot (\rho \textbf{v})=0, \\
&\frac{\partial}{\partial t}(\rho \textbf{v}) + \nabla \cdot \textbf{P}=0, \\
&\frac{\partial}{\partial t}\textbf{P} + \nabla \cdot {\Phi}_3=
\lambda_{PP}:\bm{\Theta}^{-1},
\end{align*}
where $(\Phi_3)_{ijk}=v_i P_{jk} + v_j P_{ik} + v_k P_{ij} - 2\rho v_iv_jv_k+2\bm{\Theta}^2:\partial\textbf{g}(\bm{\Theta})/\partial \bm{\Theta}$ is solved based on the constraint of entropy flux, here $\textbf g(\bm{\Theta})$ is an arbitrary vector function.
Above equations will reduce to the Euler equations in equilibrium, by splitting $\textbf{P}=\bm{\tau} + p\textbf{I} + \rho \textbf{v} \otimes \textbf{v}$
and requiring $\nabla \cdot [\bm{\Theta}^2:\partial\textbf{g}(\bm{\Theta})/\partial \bm{\Theta}]=0$.

There are other types of Euler equations also posses the CDF structure. Here we list a few of them.
For multidimensional Euler equations of gas dynamics with damping, Yong \cite{yong2008} observed that this system could be written into the form of CDF in Eq. \eqref{con-diss} with elements defined as
\begin{align*}
\textbf{U}=\begin{pmatrix}\rho\\\rho \textbf{v} \end{pmatrix}, \quad \textbf{F}(\textbf{U})=\begin{pmatrix} \rho \textbf{v}\\ \rho \textbf{v} \otimes \textbf{v} +p(\rho) \textbf{I} \end{pmatrix}, \quad
\textbf{Q}(\textbf{U})=\begin{pmatrix}0\\ -\rho \textbf{v} \end{pmatrix},
\end{align*}
where $(\rho, \rho \textbf{v})^T$ denote the respective mass density and momentum, $\textbf{I}$ is a unit matrix.
Choosing the entropy function
\begin{equation}
s(\textbf{U})=\frac{1}{2}\rho |\textbf{v}|^2 + \int^{\rho}\int^{\tau} \frac{p'(\sigma)}{\sigma} d\sigma d\tau,
\end{equation}
and specifying the dissipation matrix $\textbf{M}=\rho \textbf{I}$~($\rho>0$), one can verify the aforementioned conditions in \eqref{condition 1}-\eqref{condition 2}.

Analogously, the one-dimensional Euler equations of gas dynamics in vibrational nonequilibrium, and of viscoelastic materials in isothermal case were also included into the framework of CDF \cite{yong2008}.

\subsection{Optics, radiation and \textit{etc.}} \label{example3}
In areas of classical electromagnetism and optics, the Maxwell equation plays a fundamental
role in studying the generation and propagation of electric and magnetic fields.
A three dimensional quasi-linear evolutionary system for nonlinear optics has been shown to possess the elegant form of CDF \cite{yong2008}, which reads
\begin{equation}\label{maxwell}
\begin{split}
&\frac{\partial \textbf{D}}{\partial t} - \nabla \times \textbf{B}= 0, \\
&\frac{\partial \textbf{B}}{\partial t} + \nabla \times \textbf{E}= 0, \\
&\frac{\partial \chi}{\partial t} = |\textbf{E}|^2 - \chi,
\end{split}
\end{equation}
where the vectors $(\textbf{D}, \textbf{B}, \textbf{E})$ denote the fields of displacement, magnetic, and electric respectively, and $\textbf{D}=(1+\chi)\textbf{E}$ with $\chi>0$. Taking the state variables as $\textbf{U}=(\textbf{D}, \textbf{B}, \chi)^T$, and applying the entropy as
\begin{equation}
s(\textbf{U})=(1+\chi)^{-1}|\textbf{D}|^2+|\textbf{B}|^2+\frac{\chi^2}{2},
\end{equation}
the quasi-linear Maxwell equation in \eqref{maxwell} is readily cast into CDF in a form of \eqref{con-diss}.
Moreover, conditions in (\ref{condition 1})-(\ref{condition 2}) are also guaranteed.

Next, we move to the hydrodynamic system of Euler equations coupling with radiation transports. This system was first presented in Ref. \cite{yong2008} in detail, and was shown to be a special case of CDF. The radiation hydrodynamics could be put into the form of system \eqref{con-diss} by specifying
\begin{align}\label{radiation}
\textbf{U}=\begin{pmatrix}\rho\\ \rho v_1 \\ \rho v_2 \\ \rho v_3 \\ \rho E \\ I_1 \\ \vdots \\ I_L    \end{pmatrix}, \quad
F_j(\textbf{U})=\begin{pmatrix} \rho v_j\\ \rho v_1 v_j + \delta_{1j}p \\ \rho v_2 v_j + \delta_{2j}p \\ \rho v_3 v_j + \delta_{3j}p \\ \rho E v_j + p v_j \\ \mu_{j}^1 I_1 \\ \vdots \\ \mu_{j}^L I_L \end{pmatrix}, \quad
\textbf{Q}(\textbf{U})=\begin{pmatrix}0\\0\\0\\0\\ C\rho \sum_{l=1}^L(I_l - B(\theta))\\ -\rho (I_l - B(\theta))\\ \vdots\\ -\rho (I_L - B(\theta))   \end{pmatrix},
\end{align}
where the subscript $j$ denotes the component in the $j$-th direction ($j=1,2,3$), $I_{l}$ denotes the intensity of radiation along the direction $\mu^{l}=(\mu_1^l, \mu_2^l, \mu_3^l)$ ($l=1,2,\cdots,L$), $B(\theta)$ is the Planck function of the nonequilibrium temperature $\theta$ ($\theta \geq \theta_0$), where $\theta_0$ is a critical temperature below which assumptions of radiations fail, the coefficient $C>0$.

Since the Planck function $B(\theta)$ is strictly increasing w.r.t the temperature $\theta$, its inverse function could be solved as $\theta=b(B(\theta))$ for $\theta \geq \theta_0$. Then the entropy function for system \eqref{radiation} is defined as
\begin{equation}
s=-\rho s_0(\rho, e) - C\sum_{l=1}^L \int_{B(\theta_0)}^{I_l} \frac{dy}{b(y)},
\end{equation}
where $s_0(\rho, e)$ is the specific entropy for equations of the Euler part.

In the last example, we move to multi-component reactive flows \cite{yong2008}, which has been widely applied in  chemical engineering.
Since both hydrodynamic flows and chemical reactions have been considered before, the only unsolved difficulty is how to deal with the coupling. Consider the simplest case by neglecting the external fields, velocity of mass diffusion, heat conduction, and viscosity, then we have
\begin{align}\label{reactive}
\textbf{U}=\begin{pmatrix}\rho_1 \\ \rho_2 \\ \vdots \\ \rho_N \\ \rho v_1 \\ \rho v_2 \\ \rho v_3 \\ \rho E \end{pmatrix}, \quad
F_j(\textbf{U})=\begin{pmatrix} \rho_1 v_j\\ \rho_2 v_j \\ \vdots \\ \rho_N v_j \\ \rho v_1 v_j + \delta_{1j}p \\ \rho v_2 v_j + \delta_{2j}p \\ \rho v_3 v_j + \delta_{3j}p \\ \rho E v_j + p v_j \end{pmatrix}, \quad
\textbf{Q}(\textbf{U})=\begin{pmatrix} m_1\tau_1 \\m_2\tau_2\\\vdots\\m_N\tau_N\\ 0 \\ 0 \\ 0 \\ 0 \end{pmatrix},
\end{align}
where $\rho_i, m_i$ and $\tau_i$ denote the density, molar mass, and molar production rate of the $i$-th species ($i=1,2,\cdots,N$). The pressure is given as $p=\theta \sum_{i} \frac{R_g}{m_i} \rho_i$, with $R_g$ representing the gas constant. The total mass density is $\rho=\sum_{i} \rho_i$, and the total energy is $\rho E=\sum_{i} \rho_i [\epsilon_i^0 + \int_{\theta_0}^{\theta} c_{vi}(y) dy] + \frac{\rho}{2}|\textbf{v}|^2$, with $\epsilon_i^0=\epsilon_i(\theta_0), c_{vi}=c_{vi}(\theta)$ denoting the specific internal energy at the temperature $\theta_0$ and specific heat per unit volume, respectively.

The entropy function for reactive flows is
\begin{equation}\label{reactive entropy}
s(\textbf{U})=-\sum_{i=1}^N \rho_i s_i(\rho_i, \theta),
\quad
s_i(\rho_i, \theta)=s_i^0 + \int_{\theta_0}^{\theta} \frac{c_{vi}(y)}{y} dy - r_i \ln (\frac{\rho_i}{m_i}),
\end{equation}
where the constant $s_i^0$ is a reference entropy. In a well-stirred system, an explicit form of the production rate is given by the mass-action law as $(\tau_1, \tau_2, \cdots, \tau_N)=\frac{d}{dt}(c_1, c_2, \cdots, c_N)$, where $dc_i/dt$ is expressed in \eqref{mass-action equation}. With the relations \eqref{reactive}, \eqref{reactive entropy} and \eqref{mass-action equation} in hand, one could directly verify the conditions in \eqref{condition 1} and \eqref{condition 2}.

\section{Novel Applications}
\label{applications II}
This part goes through several novel applications of CDF relating to non-Fourier heat conduction, non-Newtonian viscoelastic fluids, wave transportation in neuroscience, soft matter physics and boundary control problem with an effort to sketch the backgrounds of these diverse fields and to present basic results derived from CDF.
Remark that the applications treated here cover a broad range of physical systems, so only main ideas closely relating to the models are stated for simplicity. Interested readers may consult references therein for further details.

\subsection{Non-Fourier Heat conduction}
As a first example, we apply the CDF to study heat conduction in rigid bodies. Here only the thermal process is taken into consideration, while the mechanical motion is neglected. As a result, the energy conservation law is written as
\begin{equation}\label{heat_conduction}
\frac{\partial u}{\partial t}+\nabla\cdot \textbf{q}=0,
\end{equation}
where $(u,\textbf{q})$ denote the internal energy and heat flux respectively. To close the heat conduction equation \eqref{heat_conduction}, we have to find the governing equation of heat flux $\textbf{q}$, which is traditionally completed by the Fourier's law.
Fourier's law provides an intuitive description of how heat flows from hot materials to cold materials. However, it is criticized for possessing an infinite speed of signal propagations and an absence of thermal fronts. Much efforts have been made to amend this problem, including the Cattaneo's law, ballistic heat propagation and many others \cite{jou1999,Krivtsov2018One}.

The generalization of Fourier's law based on CDF was first proposed by Zhu \textit{et al.} \cite{zhu2015conservation}. We revisit it here to emphasis the choice of entropy function and nonlinear constitutive relations. It is notable that, EIT adds the heat flux $\textbf{q}$ into the space of state variables directly, while CDF uses an alternative non-equilibrium (dissipative) variable $\textbf{w}$ instead. The entropy is assumed to be a strictly concave function of both the conserved and dissipative variables, $s=s(u,\textbf{w})$. The non-equilibrium temperature is defined as the partial derivative of entropy w.r.t. the internal energy, that is,
\begin{equation}
\theta^{-1}=\frac{\partial s}{\partial u}(u,\textbf{w}).
\end{equation}
By the generalized Gibbs relation, and by separating the entropy flux and entropy production rate, we obtain the evolution equation for $\textbf{w}$, and thereby for $\textbf{q}=\frac{\partial s}{\partial \textbf{w}}$ as follows,
\begin{equation}\label{generalized Fourier}
\frac{\partial \textbf{w}}{\partial t} + \nabla \theta^{-1}= \textbf{M} \cdot \textbf{q},
\end{equation}
where the dissipation matrix $\textbf{M}(u,\textbf{w})$ is positive definite. Eq. \eqref{generalized Fourier} is called the generalized (or extended) law of heat conduction. Remark that, by choosing a conjugate variable of $\textbf{q}$ and a dissipative matrix $\textbf{M}$ depending on $(u,\textbf{w})$, the formulation of CDF is truly nonlinear.

{\bf\em Cattaneo's Law.}
To illustrate the connection between generalized heat conduction equation in \eqref{generalized Fourier} with classical ones,  we specify the entropy of a quadratic type w.r.t. dissipative variables $\textbf{w}$, and a scalar diagonal dissipation matrix
\begin{equation}\label{pair}
s=s_0(u) - \frac{1}{2\alpha} |\textbf{w}|^2, \quad
\textbf{M}=\frac{1}{\lambda \theta^2}\textbf{I},
\end{equation}
where $s_0(u)$ is the equilibrium entropy, $\alpha=\alpha(u)$ is the thermal relaxation time for heat conduction, the parameter $\lambda \geq 0$.
Direct substitution of $s$ and $\textbf{M}$ into the evolution equation \eqref{generalized Fourier} yields the
generalized Cattaneo's law,
\begin{equation}\label{generalized Cattaneo' law}
\frac{\partial}{\partial t}(\alpha \textbf{q}) - \nabla\theta^{-1}=- \frac{1}{\lambda \theta^2} \textbf{q}.
\end{equation}
When $\alpha=\alpha_0$ is a constant independent on time $t$, we have $\theta^{-1}=\partial s/\partial u= T^{-1}$, then
\begin{equation}\label{classical Cattaneo' law}
\tau_0\textbf{q}_t + \lambda\nabla T + \textbf{q}=0,
\end{equation}
here $\tau_0=\alpha_0 \lambda T^2$ is the relaxation time. This is exactly the classical Cattaneo's law.
As the relaxation time $\alpha_0 \rightarrow 0$, by performing the Maxwellian iteration for $\textbf{q}$ in the generalized Cattaneo's law, we recover the well-known Fourier's law $\textbf{q}=-\lambda \nabla \theta$.

{\bf\em Thermomass Model.}
An unexpected corollary of the generalized Cattaneo's law
is that Eq. \eqref{generalized Cattaneo' law} enjoys a similar form with the thermomass model presented by Guo et al. \cite{Dong2011Generalized}.
In this case, the pair of entropy and dissipation matrix $(s,\textbf{M})$ are the same as Eq. \eqref{pair} with parameters being chosen as
\begin{equation}
\alpha(u)=\frac{\rho c_v}{2\gamma u^3}=\frac{\rho}{2\gamma c_v^2 T^3}, \quad
\textbf{M}=\frac{1}{\lambda T^2}\textbf{I},
\end{equation}
where $\rho$ denotes the density of the conduction material, $c_v$ is the heat capacity
at constant volume, $\gamma$ is the heat capacity ratio.
Substituting above parameters into the generalized Cattaneo's law, and noticing the non-equilibrium temperature $\theta^{-1}=T^{-1} + \frac{\alpha'(u)}{2\alpha^2(u)} |\textbf{c}|^2$, one has \cite{huo17}
\begin{equation}
\tau_{TM} \textbf{q}_t - 3 c_v \textbf{L} T_t + 3 (\nabla \textbf{q})^T \cdot \textbf{L}
+ \lambda(1-6M_{H}^2) \nabla T + \textbf{q}=0,
\end{equation}
where
\begin{equation*}
\tau_{TM}=\frac{\lambda \rho}{2 \gamma c_v^2 T  } , \quad
\textbf{L}=\frac{\lambda \rho}{2 \gamma c_v^3 T^2 } \textbf{q} , \quad
M_{H}^2=\frac{\lambda \rho |\textbf{q}|^2}{2 \gamma c_v^3 T^2}.
\end{equation*}
The only difference between the original thermomass model \cite{Dong2011Generalized} and the present one is the constants before $T_t$, $\nabla \textbf{q}$ and $\nabla T$.

Considering the anisotropy of heat conduction in rigid bodies or viscoelastic materials, Guyer and Krumhansl \cite{Guyer1966Solution} developed a tensorial theory, the Guyer-Krumhansl model. Huo \cite{huo17} presented that, the Guyer-Krumhansl type model could also be recovered from the generalized equations based on CDF by following the same procedures listed for Cattaneo's law.
The same argument applies to the ballistic-diffusive model \cite{Chen2001Ballistic} of heat conduction in nano-scale materials too.

\subsection{Waves transportation in neuroscience}
The axonal transport plays a key role in signal transmission of neurons.
Since the typical ratio of axon length to its diameter exceeds $1000$, we assume the transport is 1-dimensional.
Denote the concentration of the $i$-th subpopulation by $c_i=c_i(x,t)$ ($i=1,2,\cdots, N$), with $x \in \mathbb{R}^1$ being the distance from the cell body to the axon.
Based on the experimental observations and the mass-action law, one can construct the mathematical model for axonal transport, which is known as the reaction-hyperbolic system \cite{Carr1995global}.
In the form of Eq. \eqref{con-diss}, it reads
\begin{align}\label{reaction-hyperbolic}
\textbf{U}=\begin{pmatrix}c_1 \\ c_2 \\ \vdots \\ c_N \end{pmatrix}, \quad
F(\textbf{U})
=\bm{\Lambda} \textbf{U}
=\begin{pmatrix} \lambda_1 c_1\\ \lambda_2 c_2 \\ \vdots \\ \lambda_N c_N \end{pmatrix}, \quad
\textbf{Q}(\textbf{U})=\frac{1}{\varepsilon} \begin{pmatrix} f_1 \\ \vdots \\ f_i - f_{i-1}\\\vdots \\ -f_{N-1}\end{pmatrix},
\end{align}
where the constant $\lambda_i$ is the velocity of transport, $\bm{\Lambda}=diag(\lambda_1, \cdots, \lambda_N)$, and $f_i=f_i(c_i, c_{i+1})$.
Notice that the source term is stiff with a small parameter $\varepsilon >0$,  which results from the fact that the chemical reactions are much faster than transport.
Yan and Yong \cite{Yan2011Weak} proved the global existence of entropy solutions to system \eqref{reaction-hyperbolic}, and further justified the zero-relaxation limit from above system to the equilibrium system as $\varepsilon\rightarrow0$.

When the source term can be approximated by a linear combination of concentrations, Eq. \eqref{reaction-hyperbolic} reduces to the linear case with
\begin{equation}\label{linear-reaction-hyperbolic}
\textbf{Q}(\textbf{U})=\textbf{K}\textbf{U}, \quad
\textbf{K}=(k_{ij}),
\end{equation}
This model can be used to explain the approximate traveling waves observed in experiments \cite{Reed1990Approximate}.
Denote the initial and boundary conditions as
\begin{equation}\label{initial-boundary}
\textbf{U}|_{t=0}=\textbf{U}_{0}(x), \quad
\textbf{U}|_{x=0}=\textbf{U}_{0}(0),
\end{equation}
here the constraint $\bm{\Lambda}\textbf{U}_{0x}(0)=\textbf{K}\textbf{U}_{0}(0)$ is used to guarantee the continuously differentiability for solutions of the initial-boundary problem \eqref{linear-reaction-hyperbolic}-\eqref{initial-boundary}.

Denote the time-independent solution of \eqref{linear-reaction-hyperbolic}-\eqref{initial-boundary} by $\textbf{B}(x)$, \textit{i.e.}, $\bm{\Lambda} \textbf{B}_x=\textbf{K}\textbf{B}$.
Then the steady state $\textbf{B}(x)$ can be exponentially formulated as
$\textbf{B}(x)=\bm{\Lambda}^{-1} \exp(\textbf{K}\bm{\Lambda}^{-1}x)\bm{\Lambda}\textbf{U}_{0}(0)$. In general, the solution $\textbf{U}(x,t)$ of the initial-boundary-value problem \eqref{linear-reaction-hyperbolic}-\eqref{initial-boundary}
may not converge to the steady state $\textbf{B}(x)$ when the time goes to infinity. One should put suitable restrictions to guarantee the convergence.
Yan and Yong \cite{yan2012} rigorously proved the time-asymptotic stability of steady solutions by assuming the following structural conditions:
\begin{itemize}
 \item 
     $k_{ij}\geq0,\forall i\neq j$;
 \item $\sum_{i=1}^{N}k_{ij}=0, \forall j=1,2,\cdots,N$;
 \item $\textbf{K}$ is irreducibility;
 \item There exit $i$ and $j$ such that $\lambda_i\neq \lambda_j$.
\end{itemize}

\begin{theorem}
Under above structural assumptions, if $\textbf{U}_{0}(x) - \textbf{B}(x) \in H^2$,
then the linear reaction-hyperbolic system \eqref{linear-reaction-hyperbolic}-\eqref{initial-boundary} has a unique global solution $\textbf{U} \in C(0,+\infty; H^2)$ satisfying
\begin{equation*}
\lim_{t\rightarrow \infty} \sup_{x\geq 0 } |\textbf{U}(x,t) - \textbf{B}(x)|=0.
\end{equation*}
\end{theorem}

\subsection{Soft matter physics}

\subsubsection{Polymer diffusion}
Soft matters, for instance colloids, polymers and liquid crystals, are ubiquitous in nature and industry. They go beyond the conventional category of solids and fluids \cite{doi2013soft}. A common characteristics of soft matters is that they are made of large structural elements, and therefore show large, nonlinear and slow responses \cite{doi2013soft}.
Soft matter physics focuses on the structure, fluctuation, deformation, diffusion, phase transition of these systems.

As a simple and typical application of CDF in soft matter physics, we have considered the diffusion of Brownian particles in dilute solutions \cite{Peng2018Conservation1}. The particle density $n$ satisfies the conservation law of mass,
\begin{equation}\label{diffcon}
\frac{\partial n}{\partial t}+\nabla \cdot (n\textbf{v}) =0,
\end{equation}
where the average velocity $\textbf{v}(\textbf{x}, t)$ of particles is non-conserved due to friction. In the isothermal case, the free energy function is constructed as
\begin{equation}
f(t)=\frac{1}{2}n|\textbf{v}|^2+nU(\textbf{x})+k_BTn\ln n,
\end{equation}
where $\frac{1}{2} n|\textbf{v}|^2$, $nU(\textbf{x})$ and $(-k_B n\ln n)$ represent separately, the kinetic energy, the potential energy, and the mixing entropy at constant temperature. A direct calculation of the time changes of free energy gives the constitutive equation for velocities
\begin{equation}\label{diffusion-momentum}
\frac{\partial}{\partial t} (n\textbf{v})+\nabla\cdot (n \textbf{v}\otimes \textbf{v})=-n\nabla U-k_BT\nabla n-n\zeta \textbf{v},
\end{equation}
by choosing the dissipation matrix $\textbf{M}=n/\zeta$ and by using the continuity equation in \eqref{diffcon}.
Above equation turns to be the classical momentum equation for particle motion by considering the external potential force $n \nabla U$, frictional force $n\zeta \textbf{v}$, and entropic force $k_BT\nabla n$. Furthermore, in the over-damped limit as the friction coefficient $\zeta \rightarrow \infty$, by applying the Maxwellian iteration \cite{yong2004diffusive}, we obtain that
\begin{equation*}
\begin{aligned}
\textbf{v}&=-\frac{1}{n\zeta}\left(n\nabla U+k_BT\nabla n+n\frac{\partial \textbf{v}}{\partial t}+n\textbf{v}\cdot\nabla \textbf{v}\right)
 &=-\frac{1}{n\zeta}\left(n\nabla U+k_BT\nabla n\right)+o(\zeta^{-1}),
\end{aligned}
\end{equation*}
where the leading term recovers the Smoluchowskii equation \cite{doi2011,doi2013soft}.

\subsubsection{Phase separation}
In above example, the suspension of particles and fluid is assumed to be static from the macroscopic viewpoint. Here, when the fluid is in motion, the phase separation of polymer solutions has been formulated through CDF \cite{Peng2018Conservation1}.
The mass conservation laws of polymers and solvent molecules read
\begin{equation}\label{0.1}
\left\{
             \begin{array}{lr}
  \frac{\partial \phi}{\partial t}=-\nabla \cdot (\phi \textbf{v}_p ),        & \\
  \frac{\partial (1-\phi)}{\partial t}=-\nabla \cdot \left[(1-\phi)\textbf{v}_s\right] , &
             \end{array}
\right.
\end{equation}
where $\phi$ is the volume fraction of polymers, and $(1-\phi)$ of solvent molecules. We denote $\textbf{v}_p$ and $\textbf{v}_s$ as the velocities of polymers and solvent molecules, respectively. Summing up Eq. \eqref{0.1}, one arrives at the incompressible condition
\begin{equation}\label{0.2}
\nabla \cdot \textbf{v}=0
\end{equation}
for the average velocity $\textbf{v}=\phi \textbf{v}_p+(1-\phi)\textbf{v}_s$ of solutions.
Taking the polymers into account, the mixed solution possesses elasticity and viscosity at the same time. Therefore, the conservation law of total momentum becomes
\begin{equation}\label{0.4}
   \frac{\partial \textbf{v}}{\partial t}+\textbf{v}\cdot \nabla \textbf{v}
   =\nabla \cdot (-p\textbf{I} + \bm{\tau}_e + \bm{\tau}_v),
\end{equation}
where 
the symmetric tensors $\bm{\tau}_e$ and $\bm{\tau}_v$ denote elastic and viscous stresses.
Under isothermal conditions, the free energy function is
\begin{equation}\label{0.6}
f(t)=\eta(\phi)
+\frac{\alpha_0}{2}|\nabla \phi|^2  + \frac{1}{2}b^2+ \frac{1}{2}|\textbf{v}|^2+\frac{1}{2}tr(\bm{\tau}_s),
\end{equation}
where the temperature is set to be unit for simplicity, $\alpha_0\geq 0$ is a positive constant, $b \textbf{I}$ is the bulk stress tensor arising from polymer configurations. Here the mixing entropy could be modeled by the classical Flory-Huggins theory \cite{doi2013soft}, $\eta(\phi)=\frac{1}{m_p}\phi \ln \phi + \frac{1}{m_s}(1-\phi)\ln(1-\phi)+\chi \phi(1-\phi)$ with $m_p$ and $m_s$ denoting  molecular weights of polymers and solvent molecules separately, and $\chi$ characterizes the effective Flory interaction. $\bm{\tau}_s=\bm{\tau}_e+\alpha_0\nabla \phi\otimes \nabla \phi$ is recognized as the shear stress.

Utilizing the generalized Gibbs relation and separating the entropy production rate from the entropy flux, we have the following constitutive relations
\begin{equation}\label{phase separation}
\left\{
             \begin{array}{lr}
             \textbf{v}_p - \textbf{v}_s=-M(\phi) \nabla \left( \frac{\partial \eta}{\partial \phi}- \alpha_0\Delta \phi-\alpha_1 b \right), &  \\
             \bm{\tau}_v= \zeta\left[\nabla \textbf{v} +(\nabla \textbf{v})^T\right], &  \\
             \frac{d b}{dt}- \alpha_1 \nabla \cdot \left[\phi(1-\phi)(\textbf{v}_p-\textbf{v}_s)\right] = -\frac{1}{\xi_1} b, &  \\
             \frac{d}{dt}\bm{\tau}_s-(\nabla \textbf{v})^T \cdot \bm{\tau}_s-\bm{\tau}_s \cdot \nabla \textbf{v} -\alpha_2 \left[\nabla \textbf{v} + (\nabla \textbf{v})^T\right] = -\frac{1}{\xi_2} \bm{\tau}_s, &
             \end{array}
\right.
\end{equation}
based on CDF.
The first relation shows that the velocity difference between polymers and solvent molecules is caused by chemical potentials from mixing, phase separation and polymer configuration. The second formula expresses the Newton's law of viscosity with $\zeta > 0$. The third and fourth relations both belong to relaxation-type equations with parameters $\xi_1, \xi_2>0$ representing typical relaxation times for polymer compressing and solution shearing.

Combining the conservation laws \eqref{0.1}-\eqref{0.4} and constitutive relations \eqref{phase separation} together, we could readily obtain a system of closed equations for phase separation in polymer solutions, which has been studied by Zhou \textit{et al.} based on a variational approach \cite{zhou2006modified}.
The only difference between our system and that in Ref. \cite{zhou2006modified} lies on the definition of osmotic pressure.

\subsubsection{Isothermal flows of liquid crystals}
The liquid crystal is an intermediate state of materials between solids and fluids. Here we focus on the nematic liquid crystal, which consists of long, thin, rod-like molecules with long axes of neighbouring molecules aligned parallel to each other roughly.
In this section, we briefly sketch the formalism of CDF for modeling the hydrodynamic flows of nematic liquid crystals \cite{Peng2018Conservation1}.

The derivation is essentially the same as that for normal fluids, while the major difference is the choice of state variables. The molecular orientation of liquid crystals is affected by both fluid flows and external fields (magnetic or electric) \cite{doi2013soft}.
To characterize the orientational preference of rod-like molecules, a direction vector $\textbf{d}\in \mathbb R^3$ is introduced, which evolves according to
\begin{equation}\label{liquid-1}
\begin{aligned}
&\frac{\partial \textbf{d}}{\partial t}+\textbf{v} \cdot \nabla \textbf{d}=\textbf{w},
\end{aligned}
\end{equation}
where $\textbf{w} \in \mathbb R^3$ is the force moment. Together with the incompressible condition in \eqref{0.2} and momentum conservation in \eqref{0.4}, they constitute the governing equations under consideration. By setting the temperature to be unit, the free energy function is specified as
\begin{equation} \label{liquid crystal free energy}
f=\frac{1}{2}|\textbf{v}|^2+\frac{\lambda}{2}|\nabla \textbf{d}|^2+\lambda \varPhi(\textbf{d})+\frac{\gamma}{2}|\bm{\tau}_v|^2 + \frac{\gamma}{2}|\textbf{w}|^2,
\end{equation}
with the coefficients $\lambda, \gamma>0$, $\lambda$ is the ratio between kinetic energy and potential energy. Here $\varPhi(\textbf{d})=\frac{1}{2\epsilon^2}(|\textbf{d}|^2 -1 )^2$ serves as a penalty function for the constraint \cite{Lin2000Existence} on the unit length of director ($|\textbf{d}|=1$).
Adopting a diagonal dissipation matrix, CDF suggests following constitutive equations for $\bm{\tau}_e, \bm{\tau}_v$ and $\textbf{w}$:
\begin{equation}\label{constitutive}
\left\{
\begin{aligned}
&\bm{\tau}_e(\textbf{d})=-  \lambda \nabla \textbf{d} \cdot (\nabla \textbf{d})^T,\\
& \gamma \frac{\textbf{d} \bm{\tau_v}}{dt}- \nabla \textbf{v}= -\frac{1}{\alpha} \bm{\tau}_v, \\
& \gamma \frac{d \textbf{w}}{dt} + \lambda[ \varphi(\textbf{d}) -  \Delta \textbf{d}]= -\frac{1}{\beta} \textbf{w},
\end{aligned}
\right.
\end{equation}
where the elastic stress $\bm{\tau}_e$ makes no contribution to entropy production rate, the parameters $\alpha, \beta>0$ represent typical relaxation times for the viscous stress and force moment respectively. In the limit of $\alpha, \beta \rightarrow 0$, we could recover the simplified Ericksen-Leslie equations for liquid crystals \cite{Lin1995Nonparabolic} by applying Maxwellian iterations on Eq. \eqref{constitutive}.

\subsubsection{Non-isothermal flows of liquid crystals}
The modeling of non-isothermal processes is a challenging problem for most existing non-equilibrium theories. In this section, we are going to show that CDF provides a systematic way to solve the issues relating with non-isothermal situations, by taking the flow of nematic liquid crystals as a typical example \cite{Peng2018Conservation2}.

Neglecting external electric and magnetic fields, the conservation laws of mass, momentum, angular momentum and total energy for nematic liquid crystals read
\begin{align}
&\frac{\partial}{\partial t}\rho + \nabla \cdot (\rho \textbf{v})=0,\label{mass1}\\
&\frac{\partial}{\partial t}(\rho \textbf{v})+ \nabla\cdot(\rho \textbf{v} \otimes \textbf{v})=\bm{\xi} + \nabla \cdot \bm{\sigma},\label{linmom1}\\
&\frac{\partial}{\partial t}(\rho_1 \textbf{w})+ \nabla\cdot(\rho_1 \textbf{v} \otimes \textbf{w})= \textbf{g} + \nabla \cdot \bm{\pi}, \label{angmom1}\\
&\frac{\partial}{\partial t}(\rho e) + \nabla \cdot(\rho \textbf{v} e)=\bm{\xi} \cdot \textbf{v} + \nabla \cdot (\bm{\sigma} \cdot \textbf{v} + \bm{\pi} \cdot \textbf{w} - \textbf{q})\label{energy1},
\end{align}
where $\rho$ is the density, $\bm{\xi}$ is the external body force per unit volume, $\bm{\sigma}$ is the stress tensor. In Eq. \eqref{angmom1}, $\rho_1=\rho |\textbf{r}|^2$ is the density of inertia moment,
with $\textbf{r}$ denoting the effective position vector and its norm $|\textbf{r}|$ is assumed to be constant. The material derivative of director vector gives the director velocity $\textbf{w} \equiv \frac{d}{dt}(\textbf{d})$.
$\bm{\pi}$ is the director surface torque, $\textbf{g}$ is the intrinsic body torque. Both of them describes the influence of macroscopic flows on the microscopic structure.
In Eq. \eqref{energy1}, the specific total energy density $e=\frac{1}{2}|\textbf{v}|^2+ \frac{\rho_1}{2\rho}|\textbf{w}|^2+u$ includes both translational and rotational kinetic energies and the internal energy $u$, $\textbf{q}$ is the heat flux.

\begin{remark}
In many previous studies, the left-hand side of Eq. \eqref{angmom1} has been neglected.
However, this inertial term plays a key role when the anisotropic axis is subjected to large accelerations \cite{Leslie1979Theory,Stewart2004The}.
Moreover, the director vector is usually simplified to be unit.
We introduce $\textbf{d}\in \mathbb {R}^3$ as a 3-dimensional vector, to account for the preferred orientation and the average length \cite{Flory1984Molecular} of rod-like molecules.
In this way, the vectorial theory could possibly describe the fluids of mixtures of molecules with varied lengths \cite{Lekkerkerker1984On, Flory1984Molecular, He2016Isotropic}.
\end{remark}

Notice that, in the above model, the stress tensor is directly separated into elastic and viscous stresses as well as the thermodynamic pressure. Here we further decompose the viscous stress tensor into two parts based on their different origins,
\begin{equation}\label{sigma_EL}
\bm{\sigma}= -p\textbf{I} + \bm{\sigma}_E+ (\bm{\sigma}_V+\bm{\sigma}_L),
\end{equation}
in which $\bm{\sigma}_V$ is the viscous stress for homogenous fluid flows, and $\bm{\sigma}_L$ is the orientation-induced viscous stress. Similar decompositions hold for the director surface torque $\bm{\pi}$ and body torque $\textbf{g}$ as
\begin{equation}
\bm{\pi}=\bm{\pi}_V+\bm{\pi}_E+\bm{\pi}_L,\quad \textbf{g}=\textbf{g}_V+\textbf{g}_E+\textbf{g}_L,
\end{equation}
with the subscript $V, E, L$ denoting the homogenous viscous part, Ericksen part for
the static state, and Leslie part for the non-equilibrium state, respectively.

Now we introduce a strictly concave mathematical entropy function
\begin{equation}\label{entropy}
\eta = \rho s(\nu, u, \textbf{d}, \nabla \textbf{d}, \textbf{C}, \textbf{K},\textbf{l},\textbf{h}),
\end{equation}
where $\nu=1/\rho$, $(\textbf{C}, \textbf{K})$ are tensors with the same size of $(\bm{\sigma},\bm{\pi})$, and $(\textbf{l},\textbf{h})$ are vectors with the same size of $(\textbf{g},\textbf{q})$. $(\textbf{C}, \textbf{K}, \textbf{l})$ are used to describe the viscous-elastic effects of nematic liquid crystal flows, and $\textbf{h}$ characterizes the heat conduction induced by temperature gradients.
Therefore, the non-equilibrium temperature $\theta$ and thermodynamic pressure $p$ are defined by
\begin{equation}
\theta^{-1}=\frac{\partial s}{\partial u}, \quad \theta^{-1}p=\frac{\partial s}{\partial \nu}.
\end{equation}
The time evolution equations are constructed based on CDF, with choices of entropy in \eqref{entropy} and a diagonal and constant dissipation matrix as
$\frac{1}{\theta}diag(\frac{1}{\gamma_1}, \frac{1}{\gamma_2}, \frac{1}{\gamma_3}, \frac{1}{\theta \gamma_4})$ (coefficients $\gamma_i >0$). Thereby, the deduced constitutive relations read
\begin{equation}\label{nonisothermal EL}
\left\{
\begin{aligned}
&  (\rho \textbf{C})_t+\nabla\cdot(\rho \textbf{v} \otimes \textbf{C}) + \theta^{-1}\textbf{A}
= - \frac{1}{\theta \gamma_1}\frac{\partial s}{\partial \textbf{C}},\\
&  (\rho \textbf{K})_t+\nabla\cdot(\rho \textbf{v}\otimes \textbf{K}) + \theta^{-1}\textbf{M}
= - \frac{1}{\theta \gamma_2}\frac{\partial s}{\partial \textbf{K}}, \\
&  (\rho \textbf{l})_t+\nabla\cdot(\rho \textbf{v} \otimes \textbf{l}) -\theta^{-1} \textbf{N}
= - \frac{1}{\theta \gamma_3}\frac{\partial s}{\partial \textbf{l}},\\
&  (\rho \textbf{h})_t+\nabla\cdot(\rho \textbf{v}\otimes \textbf{h}) +\nabla \theta^{-1}
= - \frac{1}{{\theta}^2 \gamma_4}\frac{\partial s}{\partial \textbf{h}},\\
& \bm{\sigma}_E=  \rho \theta (\nabla \textbf{d} \cdot \frac{\partial s}{\partial \nabla \textbf{d}})^T,\quad
\bm{\pi}_E   = -\rho \theta (\frac{\partial s}{\partial \nabla \textbf{d}})^T,\quad
\textbf{g}_E     =  \rho \theta \frac{\partial s}{\partial \textbf{d}}, \\
&  \bm{\sigma}_L =\alpha_1 (\textbf{d}^T \cdot \textbf{A} \cdot \textbf{d}) \textbf{d} \otimes \textbf{d}  + \alpha_2 \textbf{N} \otimes \textbf{d}  +\alpha_3 \textbf{d} \otimes \textbf{N}
+ \alpha_4 \textbf{A}  \\
&~~~~~~ +\alpha_5 \textbf{d} \otimes  (\textbf{A} \cdot \textbf{d})  +\alpha_6 (\textbf{A} \cdot \textbf{d}) \otimes \textbf{d}, \\
&\bm{\pi}_L=0,\quad
\textbf{g}_L=(\alpha_2-\alpha_3)\textbf{N} + (\alpha_5- \alpha_6)\textbf{A} \cdot \textbf{d},
\end{aligned}
\right.
\end{equation}
where $2A_{ij}=v_{i,j}+ v_{j,i}, ~2 \Omega_{ij}=v_{i,j} - v_{j,i},~N_i=w_i- \Omega_{ik} d_k,~  M_{ij}= w_{j,i}+ \Omega_{kj} d_{k,i}$. Here we adopt the principle of material frame-indifference of state variables $(\textbf{A}, \textbf{M}, \textbf{N})$. $\textbf{A}$ is the symmetric part of velocity gradient, $\textbf{N}$ is the relative angular velocity measured by an observer whose reference is carried and rotated with fluids, $\textbf{M}$ is the corresponding objective variable of the gradient of angular velocity.

Following the Oseen-Frank elastic energy \cite{Stewart2004The,Frank1958I}, the classical isothermal Ericksen-Leslie model is formally shown to be a special case of our new vectorial model, in the limit of isothermal, incompressible and stationary condition. Under non-isothermal conditions, the vectorial model in Eqs. \eqref{mass1}-\eqref{energy1} and \eqref{nonisothermal EL} for flows of nematic liquid crystals could be generalized to the tensorial case, which recovers to Qian-Sheng model \cite{Qian1998Generalized} in the isothermal limit.

\subsection{Boundary control of linear hyperbolic balance laws}
In previous applications, we have always assumed the state variables distribute in the entire space, $\textbf{U}\in \mathbb{R}^{n+m}$.
However, for real systems, which are of finite size and have irregular edges, the general boundary conditions are of great interest for theoretical
modeling and numerical simulations. As an illustration, here we consider the boundary control problem for one-dimensional linear hyperbolic balance laws.
\begin{align}\label{boundary control}
\textbf{U}=\begin{pmatrix} \textbf{y}\\ \textbf{z} \end{pmatrix}, \quad
F(\textbf{U})
=\textbf{A} \textbf{U}
=\begin{pmatrix} \textbf{a} & \textbf{b}\\ \textbf{c} & \textbf{d}\end{pmatrix}
\begin{pmatrix} \textbf{y}\\ \textbf{z} \end{pmatrix}, \quad
\textbf{Q}(\textbf{U})=\begin{pmatrix} \textbf{0} \\ -\textbf{e}\textbf{z} \end{pmatrix},
\end{align}
where $\textbf{y}=\textbf{y}(t,{x}) \in \mathbb{R}^n$ and $\textbf{z}=\textbf{z}(t,{x}) \in \mathbb{R}^m$ depend on $t \geq 0$ and ${x} \in [0,1] \subset \mathbb{R}^1$. In particular, we focus on the exponential stability of the steady states, which is crucial for engineering, such as the transportation of electricity, fluid flow in open channels and road traffic \cite{Bastin2016stability}. It ensures that, the time trajectories of the system will exponentially converge to its steady states, starting from any given initial conditions.

{\bf\em Non-characteristic boundary.}
As to this system, we first assume the matrix $\textbf{A}$ has no vanishing eigenvalues. The case of vanishing eigenvalues (or equivalently, zero characteristic speeds) will be treated later. Herty and Yong \cite{Herty2016feedback, Yong2019boundary} used CDF to derive new stabilization results, by making the following structural assumptions:
\begin{itemize}
 \item (A1) There exists a symmetric positive-definite matrix
 $\textbf{A}_{0}\in \mathbb{R}^{n+m}$ such that $\textbf{A}_{0}\textbf{A}$
 is symmetric and $\textbf{A}_{0}=
     \begin{pmatrix}
     \textbf{X}_1 & 0\\ 0 & \textbf{X}_2
     \end{pmatrix}$ is block diagonal, with $\textbf{X}_1 \in \mathbb{R}^{n\times n}, \textbf{X}_2 \in \mathbb{R}^{m\times m}$;
 \item (A2) $(\textbf{X}_2 \textbf{e} + \textbf{e}^T \textbf{X}_2)$ is positive definite.
\end{itemize}

\begin{theorem}\label{nonchara}
Under the assumptions $(A1)$ and $(A2)$,
the system \eqref{boundary control} has a feedback boundary control such that
the initial boundary value problem of \eqref{boundary control} is exponentially stable,
that is, there exist constants $\nu>0, C > 0$, such that
\begin{equation*}
\| (\textbf{y, z})(t, \cdot)  \|_{L^2((0,1); \mathbb{R}^{m+n})} \leq
C \exp(-\nu t)\| (\textbf{y}_0, \textbf{z}_0) \|_{L^2((0,1); \mathbb{R}^{m+n})},
\end{equation*}
for every $(\textbf{y}_0, \textbf{z}_0) \in L^2((0,1); \mathbb{R}^{m+n})$ with initial value $\textbf{y}_0(x)=\textbf{y}_0(0, x)$, $\textbf{z}_0(x)=\textbf{z}_0(0, x)$.
\end{theorem}

A similar dissipative boundary condition has been deduced \cite{Diagne2012lyapunov} by imposing the assumption of diagonally marginally stable on the source terms.
Unfortunately, this condition can not be verified directly.
In contrast, the assumptions of (A1)-(A2) can be checked more straightforwardly and enjoy a clear physical meaning.

{\bf\em Characteristic boundary.}
Next we assume the matrix $\textbf{A}$ has zero eigenvalues.
Based on the conditions (A1)-(A2), and (A3) stated as follows:
\begin{itemize}
 \item (A3) The $m\times m$ matrix $\textbf{a}$ has only positive eigenvalues.
\end{itemize}
Yong \cite{Yong2019boundary} generalized the results from non-characteristic boundaries to characteristic
boundaries.

\begin{theorem}\label{chara}
Under the assumptions (A1), (A2) and (A3),
the system \eqref{boundary control} has a feedback boundary control such that
the initial boundary value problem of \eqref{boundary control} is exponentially stable.
\end{theorem}

Theorems \ref{nonchara} and \ref{chara} have been applied to the feedback boundary control of
water flows in open canals \cite{Herty2016feedback}, and to the transport of neurofilaments in axons \cite{Yong2019boundary}, by verifying the related structural stability conditions.
As to the hyperbolic system \eqref{boundary control} with stiff source terms (\textit{i.e.}, $\textbf{Q}(\textbf{U})/{\varepsilon}, {\varepsilon} \ll 1$), thanks to the conservation-dissipation structure, further stabilization results for the boundary control problem were derived by Herty and Yu \cite{herty2018feedback}.

{\section{Validation of CDF}}\label{validation}
To demonstrate CDF indeed provides a mathematical rigorous and physically meaningful description of irreversible processes, we state from three aspects -- mathematical analysis, numerical simulations and experimental validations based on recent advances in this direction.

\subsection{Global existence for viscoelastic fluids with finite strain}
A major advantage of CDF over other physical based modeling lays on its mathematical rigorousness, especially the well-posedness of global smooth solutions. For example, the global existence and smoothness of classical Navier-Stokes solutions is one of the seven millennium prize problems funded by Clay mathematics institute, while this problem is easily justified for our generalized NSF equations in \eqref{constitutive of concrete type}. To make a further illustration on the mathematical merit of CDF, we look into the problem of viscoelastic fluids with finite strain.

Viscoelastic fluids possess the characteristics of both viscosity from fluids and elasticity from polymers at the same time. The study of hydrodynamical models of viscoelastic fluids has a long history, which could be dated back to Maxwell \textit{et al.} \cite{Maxwell2013The_Scientific, Oldroyd1950On, deGennes1995The, Weissenberg1947A}. Combing the rational 
thermodynamics with the theory of finite strain, Coleman \textit{et al.} \cite{Coleman1961Foundations, Coleman1964Thermodynamics} put forward a class of nonlinear models for materials with finite deformations and long-time memories. However, when deriving the upper convected Maxwell model based on RT, the energy dissipation becomes negative and therefore conflicts with the second law of thermodynamics \cite{huo17}.

To overcome the difficulty, Huo \textit{et al.} \cite{huo17} developed a conservation-dissipation formalism for viscoelastic fluids with finite inelastic deformations.
In addition to the usual conservation laws of mass, momentum and total energy in Eqs. \eqref{mass}-\eqref{energy}, the finite deformation has also been taken into account.
The deformation tensor $\textbf{F}=(F_{ij})$ evolves according to
\begin{equation}\label{deform}
\frac{\partial}{\partial t}(\rho \textbf{F}) + \nabla \cdot (\rho \textbf{F} \otimes \textbf{v}) - \nabla \cdot (\rho \textbf{v} \otimes \textbf{F}^T)=0,
\end{equation}
where $F_{ij}=\partial x_i/ \partial X_j$. The vectors $\textbf x$ and $\textbf{X}$ denote the respective Eulerian (or referee) and Lagrange (or material) coordinates of the flow field.
Moreover, $\textbf{F}$ satisfies three compatibility conditions:
\begin{equation}\label{compatibility}
\nabla \cdot (\rho \textbf{F}^T)=0, \quad
F_{lj}\partial_{x_l}F_{ik}=F_{lk}\partial_{x_l}F_{ij}, \quad
\rho \det \textbf{F}=1.
\end{equation}
Together with the deformation in Eq. \eqref{deform} and Eqs. \eqref{mass}-\eqref{energy}, we have four local conservation laws for variables $(\rho, \rho \textbf{v}, \rho e, \rho \textbf{F})$. In the theory of finite strain, the stress tensor $\bm{\sigma}$ is a function of the deformation $\textbf{F}$, so that a central task of non-equilibrium thermodynamics is to find out the proper relations among these unknown variables.

According to CDF, the entropy function and dissipation matrix read
\begin{equation}\label{deform_entropy_0}
\eta=\rho s(\nu,u, \textbf{F}, \textbf{w}, \textbf{c}), \quad s=s_0(\nu,u) - \Phi(\textbf{F}) - \frac{1}{2\nu\epsilon_0}|\textbf{w}|^2 -\frac{1}{2\nu\epsilon_1}\textbf{c}:\textbf{c},
\end{equation}
and
\begin{equation}\label{deform_dissipation}
 \textbf M=\begin{pmatrix}
 \frac{1}{\theta^2 \lambda} & 0\\
  0                         & \theta(\frac{1}{\kappa} + \frac{1}{\xi} )
 \end{pmatrix},
\end{equation}
where $s$ is the specific entropy, $\nu=1/\rho$ is the specific volume, $\theta^{-1}=s_u$ is the non-equilibrium temperature, $(\textbf{w},\textbf{c})$ are conjugate variables of $(\textbf{q}, \bm{\sigma})$ respectively, the coefficients $\epsilon_0, \epsilon_1, \lambda, \kappa, \xi >0$.

\begin{remark}
Note that, the entropy function in Eq. \eqref{deform_entropy_0} is not concave w.r.t. its arguments $(\nu,u, \textbf{F}, \textbf{w}, \textbf{c})$, due to the constraints of deformation tensor and material frame-indifference. Therefore, the condition \eqref{condition 1} has to be relaxed, while the condition \eqref{condition 2} still holds \cite{huo17}.
\end{remark}

Notice that the stress tensor is given by $\bm{\sigma}=-p\bf{I}+\bm{\tau}_e + \bm{\tau}_v$, where the thermodynamic part is $p=\theta s_{\nu}$, the elastic part is $\bm{\tau}_e=\theta \rho {\Phi}_{\textbf{F}} \textbf{F}^T$, and the viscous part $\bm{\tau}_v$ is derived as follows.
Calculating the time change of entropy, and choosing the conjugate variables $s_{\textbf{w}}=\textbf{q}$,
$s_{\textbf{c}}=\theta^{-1}\bm{\tau}_v$,
we have
\begin{equation}\label{deform_constitutive}
\left\{
             \begin{array}{lr}
             \epsilon_0 [\partial_t \textbf{q} + \nabla \cdot (\textbf{q} \otimes \textbf{v})] -\nabla \theta^{-1}=- \frac{1}{\theta^2 \lambda} \textbf{q}, &  \\
             \epsilon_1 [\partial_t (\theta^{-1} \dot{\bm{\tau_v}}) + \nabla \cdot (\theta^{-1} \dot{\bm{\tau_v}} \otimes \textbf{v})] - \dot{\textbf{A}}=- \frac{1}{\kappa}\dot{\bm{\tau_v}}, &  \\
             \epsilon_1 [\partial_t (\theta^{-1} \mathring{\bm{\tau_v}}) + \nabla \cdot (\theta^{-1} \mathring{\bm{\tau_v}} \otimes \textbf{v})] - \mathring{\textbf{A}} =- \frac{1}{\xi}\mathring{\bm{\tau_v}}, &
             \end{array}
\right.
\end{equation}
based on CDF.

After substituting above equations into the conservation laws in Eqs. \eqref{mass}-\eqref{energy} and \eqref{deform}, one arrives at the Maxwell model with finite strain.
This model recovers to the one presented by Lin \textit{et al.} \cite{viscoelastic-fluids} through Maxwellian iteration when the relaxation times $\kappa, \xi$ are small enough.
Thanks to the CDF structure, a new proof on the global existence of the viscoelastic fluids with infinite Weissenberg number was addressed, both for the compressible and incompressible cases \cite{Huo2017Global}.

\begin{theorem}
(Huo \textit{et al.} \cite{Huo2017Global})
Consider our above proposed Maxwell model with finite strain under isothermal conditions. Let $\textbf{U}=(\rho,\textbf{v},\textbf{F})$ be its solution, with the equilibrium point $\textbf{U}_e=(\rho_e>0, \textbf{0}, \textbf{I}_{{\Lambda} \times {\Lambda}})$. Further suppose $\textbf{U}_0-\textbf{U}_e\in H^s$ where $s>[{\Lambda}/2]+1$ is a positive integer, $\|\textbf{U}_0-\textbf{U}_e\|_{H^s}$
is sufficiently small, and $\textbf{U}_0$ satisfies the compatibility conditions \eqref{compatibility}. Then
there exists a unique global solution $\textbf{U}=\textbf{U}(\textbf{x},t)$, with $\textbf{U}_0$ as
initial data, satisfying
\begin{eqnarray*}
&&\textbf{U}-\textbf{U}_e\in C([0,\infty),H^s)\cap L^2([0,\infty),H^s),\; \textbf{v}\in L^2([0,\infty),H^{s+1}),\;\\
&&\|\textbf{U}(T)-\textbf{U}_e\|^2_{H^s}+\int_0^T[\|\nabla \textbf{v}(t)\|^2_{H^s}+\|\nabla \textbf{U}(t)\|^2_{H^{s-1}}]dt\leq c_0\|\textbf{U}_0-\textbf{U}_e\|^2_{H^s}.
\end{eqnarray*}
\end{theorem}

\subsection{Unstable modes of BISQ model in geophysics}\label{BISQ}
CDF provides a very general framework for establishing models with long-time asymptotic stability in thermodynamics. So that if a thermodynamic model does not fall into the category of CDF structure, it may contain unstable modes violating the second law of thermodynamics and thus nonphysical. Here we illustrate such an example in geophysics.

The study of seismic waves propagating in rocks helps to locate and predict the volume of hydrocarbon. Various theories are devoted to explain the attenuation and dispersion of seismic waves. Among them, the mechanism of Biot/squirt flow (BISQ) \cite{Dvorkin1993Dynamic} stands out as a foundation to study acoustic wave propagation in saturated porous media.
However, a rigorous analysis on the stability of the BISQ model remained lacking, until the work of Liu and Yong \cite{Liu2016Stability2} was carried out.

The stability analysis is based on the hyperbolic systems of first-order PDEs:
\begin{equation}\label{con-diss1}
  \textbf{A} \partial_t \textbf{U}+ \sum_{j=1}^{\Lambda} \textbf{A}_j \partial_{x_j}\textbf{U}=\textbf{Q}(\textbf{U}),
\end{equation}
which could be regarded as a general form of system \eqref{con-diss}. $\textbf{A}$ and $\textbf{A}_j$ are matrices, and in particular $\textbf{A}$ is invertible. The fundamental stability criteria of \eqref{con-diss1} was discussed in Ref. \cite{Yong1999singular, Yong2001basic}. The original Biot equation was casted into Eq. \eqref{con-diss1}, and was proved to satisfy the stability condition, which guarantees the uniform boundness of its solutions.
However, the BISQ model, as extensions of the Biot equation with microscopic rock properties,
allows exponentially exploding solutions, as time goes to infinity.
We state the main result in the following theorem.
\begin{theorem}
(Liu \textit{et al.} \cite{liu2016stability})
The one-dimensional BISQ model has time-exponentially exploding solutions when the characteristic squit-flow coefficients is negative or has a non-zero imaginary part.
\end{theorem}

\begin{figure}[!htp]
\centering
\includegraphics[width=0.8\linewidth]{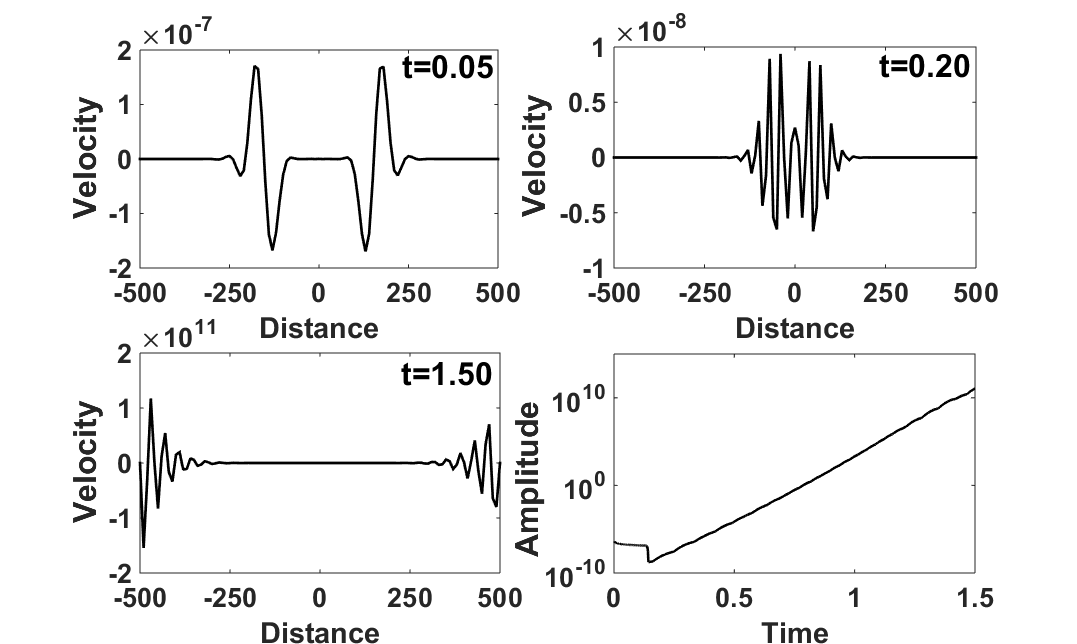}
\caption{Unstable modes of BISQ model. Adopted from Ref. \cite{liu16}.}
\label{fig.BISQ}
\end{figure}

By using second-order time splitting methods, the unstable modes of 1-d BISQ model under experimental conditions were numerically explored, with spatial distance $x \in [-500, 500]$ meters. From Fig. \ref{fig.BISQ}, at the very beginning of wave propagation, the waveform was smooth and absorbed stably by the boundary. Later, at $t=0.2s$ the propagation of the main waveform was completed, and a sub-waveform emerged hinting the oscillation of BISQ solutions. Finally, after a relatively long time, serious oscillation of BISQ solutions was clearly observed, whose amplitude grows in an explosive way.

The existence of exploding solutions along with the linearity of the BISQ model means that the BISQ model is unstable. This gives a direct explanation for the unreliability of BISQ models at low frequencies, which is well-known in experiments but has been proved theoretically for the first time.
Above analysis can be extended to the three-dimensional isotropic BISQ model.
Particularly, it was proved to be stable when the squirt-flow coefficient is positive \cite{liu2016stability}.

\subsection{Vibrations of bipyramidal particles in viscoelastic fluids}
Recently, a quite decisive hydrodynamic validation of CDF in comparison with other five classical models for compressible viscoelastic flows, including Edwards-Beris model, Oldroyd model, Oldroyd-B model, and etc., was carried out by Chakraborty and Sader \cite{Chakraborty2015nano}. According to Landau and Lifshitz \cite{Landau1987}, for compressible viscoelastic flows tending to equilibrium, the mechanical and thermodynamic pressures display a frequency dependence, \textit{i.e.}, $p_m=p-\mu_B(\nabla\cdot  \textbf{v})/(1-i\omega\lambda)$, where $\lambda$ is the relaxation time, $\omega$ is the angular frequency, $\mu_B$ is the bulk viscosity of the fluid at thermodynamic equilibrium. Chakraborty and Sader \cite{Chakraborty2015nano} pointed out that constitutive relations in Eqs. \eqref{constitutive of concrete type} constructed by CDF is ``{\em the only model that captures the correct behavior}'' of this classical thermodynamic result.

In addition, it was observed that all models display a linear dependence on the rate-of-strain tensor, but some assume a zero bulk
viscosity at low frequency. While in the high frequency limit, all other models predict ``fluid-like'' behavior with the deviatoric stress tensor remaining proportional
to the rate-of-strain. In striking contrast, our CDF model intrinsically predicts the required behavior of
an elastic solid, whose stress is proportional to and in-phase with the strain.

Chakraborty and Sader \cite{Chakraborty2015nano} further studied gold bipyramidal nanoparticles undergoing extensional mode vibrations in glycerol-water mixtures. As shown through both the
resonant frequency and quality factor in Fig. 2, numerical solutions of our CDF model by finite element methods agree quite well with the experimental data. So that they concluded that our ``{\em compressible model encompasses the general case and can be used to calculate the flows generated by arbitrarily shaped nanoscale devices immersed in simple liquids}.''

\begin{figure}[!htp]
\centering
\includegraphics[width=0.7\linewidth]{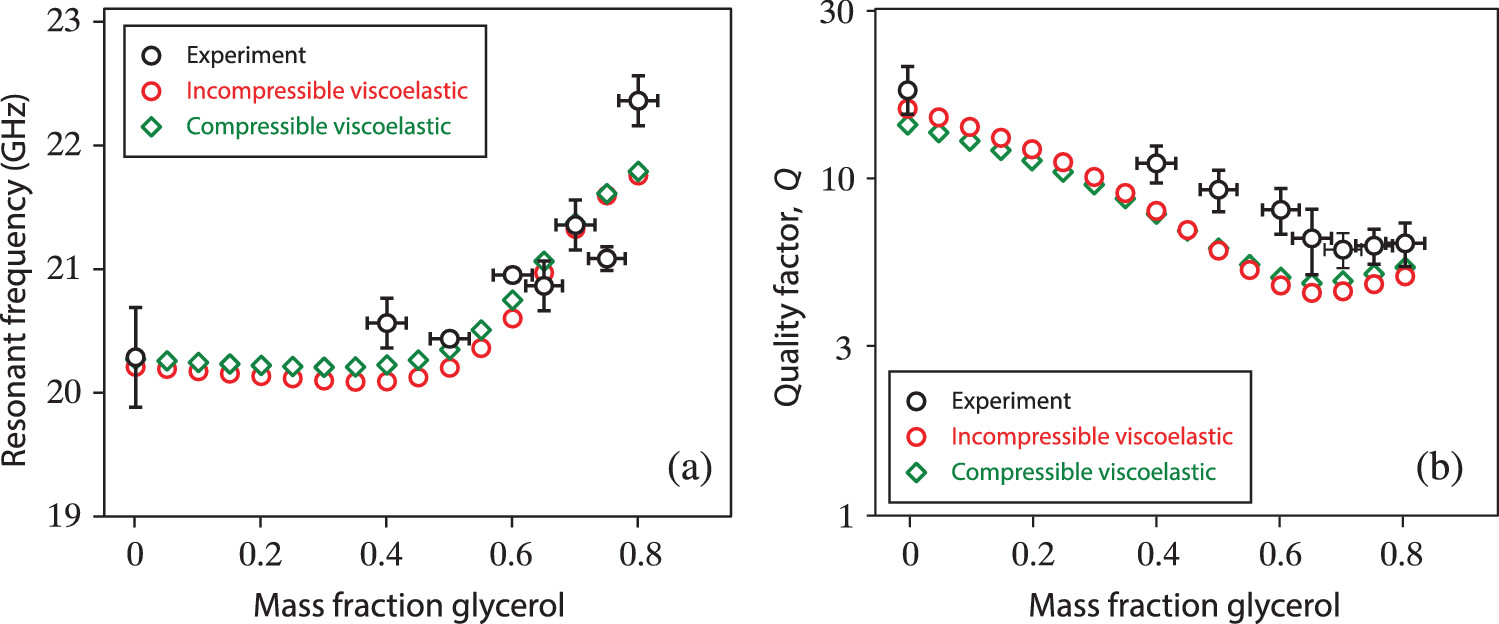}
\caption{Bipyramidal particle undergoing extensional model vibrations in glycerol-water mixtures. Adopted from Ref. \cite{Chakraborty2015nano}.}
\label{fig.nano}
\end{figure}

\section{Conclusion}
In this review, we have carefully revisited the recent advances of the Conservation-Dissipation Formalism. As a general modeling approach, CDF not only inherits the elegant mathematical structure of symmetrizable hyperbolic equations, such as the existence of a concave entropy function, the well-posedness of global smooth solutions, the asymptotic stability of long-time solutions tending to equilibrium, but also fulfills the physical requirements of non-equilibrium thermodynamics, including the mass, momentum and energy conservation in accordance with the first law of thermodynamics, a positive entropy production rate to maintain the irreversibility of underlying processes, the Onsager's relation, \textit{etc}. We notice that the physical considerations have already been widely adopted in previous thermodynamics-based approaches, like CIT, RT, EIT, GENERIC and so on, but the mathematical requirements, especially the well-posedness of solutions and their long-time asymptotic behaviors, have seldom been discussed and included into the modeling. And we believe CDF makes up for the lacking in this respect.

Rather than a useless abstract framework, CDF actually is a practical guiding principle for constructing both physically meaningful and mathematically rigorous models. Just as we have shown, not only many classical models in mathematical physics, including master equations, Fokker-Planck equations, mass-action equations and moment hierarchies of Boltzmann equations, fall into the category of CDF structure, but also diverse non-equilibrium systems in different fields, such as non-Fourier heat conduction, viscoelastic fluids, axonal transportation in neuroscience, soft matter physics, all have witnessed the successful applications of CDF. And we expect CDF can be applied to other interesting studies in the field of non-equilibrium thermodynamics in the future.

\section*{Acknowledgement}
This work was supported by the National Natural Science Foundation of China (Grants 21877070) and the Startup Research Funding of Minjiang University (mjy19033). We thank Prof. Wen-An Yong, Yi Zhu, Yucheng Hu for their stimulating discussions.

\end{document}